\documentclass[preprint]{aastex63}

\usepackage{array,multirow}
\usepackage{amsmath}
\usepackage{color,soul}
\usepackage{textcomp}
\usepackage{booktabs}
\usepackage{CJK}

\accepted{\today}
\submitjournal{The Astrophysical Journal}

\shorttitle{Tholin's surface energy}
\shortauthors{Yu et al.}

\begin{document}

\title{Surface Energy of the Titan Aerosol Analog `Tholin'}

\correspondingauthor{Xinting Yu}
\email{xintingyu@ucsc.edu}

\author[0000-0002-7479-1437]{Xinting Yu\begin{CJK*}{UTF8}{gbsn}
(余馨婷)\end{CJK*}}
\affiliation{Department of Earth and Planetary Sciences \\
University of California Santa Cruz \\
1156 High Street\\
Santa Cruz, California 95064, USA}
\affiliation{Department of Earth and Planetary Sciences \\
Johns Hopkins University \\
3400 N. Charles Street\\
Baltimore, Maryland 21218, USA}

\author{Sarah M. H\"orst}
\affiliation{Department of Earth and Planetary Sciences \\
Johns Hopkins University \\
3400 N. Charles Street\\
Baltimore, Maryland 21218, USA}

\author{Chao He}
\affiliation{Department of Earth and Planetary Sciences \\
Johns Hopkins University \\
3400 N. Charles Street\\
Baltimore, Maryland 21218, USA}

\author{Patricia McGuiggan}
\affiliation{Department of Materials Science and Engineering\\
Johns Hopkins University \\
3400 N. Charles Street\\
Baltimore, Maryland 21218, USA}

\author{Kai Kristiansen}
\affiliation{Department of Chemical Engineering \\
University of California Santa Barbara \\
Santa Barbara, California 93106, USA}

\author{Xi Zhang}
\affiliation{Department of Earth and Planetary Sciences \\
University of California Santa Cruz \\
1156 High Street\\
Santa Cruz, California 95064, USA}

\begin{abstract}

The photochemical haze produced in the upper atmosphere of Titan plays a key role in various atmospheric and surface processes on Titan. The surface energy, one important physical properties of the haze, is crucial for understanding the growth of the haze particles and can be used to predict their wetting behavior with solid and liquid species on Titan. We produced Titan analog haze materials, so-called "tholin", with different energy sources and measured their surface energies through contact angle and direct force measurements. From the contact angle measurement, we found that the tholins produced by cold plasma and UV irradiation have total surface energy around 60--70 mJ/m$^2$. The direct force measurement yields a total surface energy of $\sim$66 mJ/m$^2$ for plasma tholin. The surface energy of tholin is relatively high compared to common polymers, indicating its high cohesiveness. Therefore, the Titan haze particles would likely coagulate easily to form bigger particles, while the haze-derived surface sand particles would need higher wind speed to be mobilized because of the high interparticle cohesion. The high surface energy of tholins also makes them easily wettable by Titan's atmospheric hydrocarbon condensates and surface liquids. Thus, the hazes particles are likely good cloud condensation nuclei (CCN) for hydrocarbon clouds (methane and ethane) to nucleate and grow. And if the hazes particles are denser compared to the lake liquids, they would likely sink into the lakes instead of forming a floating film to dampen the lake surface waves.
\end{abstract}

\keywords{planets and satellites: atmospheres --- 
planets and satellites: composition --- planets and satellites: surfaces}

\section{Introduction} \label{sec:intro}

Titan is an organic rich world with photochemistry produced haze prevalent in its atmosphere and on its surface. In the Titan's upper atmosphere, N$_2$ and CH$_4$ are dissociated by UV photons and energetic particles from Saturn's magnetosphere, which leads to the formation of a variety of simple hydrocarbon and nitrile molecules such as C$_2$H$_6$, C$_2$H$_2$, HCN, C$_6$H$_6$, etc. These simple organic molecules could further polymerize and coagulate to form the larger and more complex solid organic particles, which become the refractory aerosols that form Titan's thick opaque haze layers. In Titan's atmosphere, the solid aerosol particles could serve as the cloud seeds for volatile hydrocarbon and nitrile condensates (Curtis et al., 2008; Lavvas et al., 2011) and form various types of organic clouds (H\"orst, 2017; Anderson et al., 2018). The solid haze particles in the atmosphere could also deposit onto the surface and are believed to make up portions of the surface material. For example, the spectrally dark dunes are proposed to be made up of mainly the organics produced in the atmosphere (Soderblom et al., 2007; Lopes et al. 2020). When the haze particles fall towards the polar regions of Titan, they may interact with the liquid hydrocarbon lakes. If the aerosols are able to float and form an organic film on the lake surface, it could reduce the momentum, radiation, matter, and heat transfer between the atmosphere and the seas (Cordier and Carrasco, 2019). For example, the momentum transfer could be reduced due to the aerosol film and could thus dampen the lake surface waves, which could explain Cassini's observations suggesting that waves have very low amplitudes (Stephan et al., 2010; Barnes et al., 2011; Soderblom et al., 2012; Zebker et al., 2014; Grima et al., 2017). While if the aerosols sink into lakes, they would become Titan's lakebed sediments.

\begin{figure}[h]
\centering
\includegraphics[width=32pc]{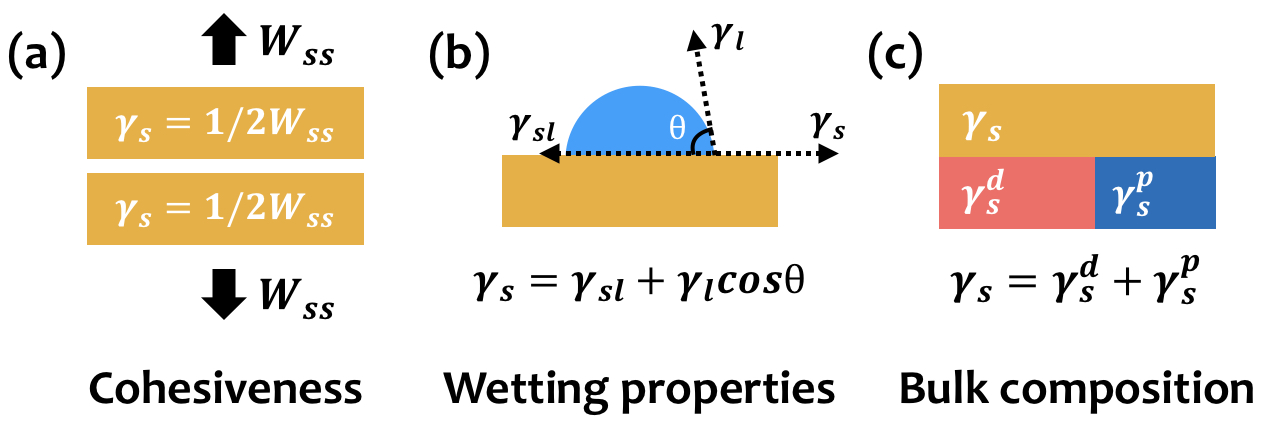}
\caption{(a) $W$ is the work needed to separate unit area of two surfaces. For two identical solid materials, it is called the work of cohesion ($W_{ss}$). The surface energy of a solid ($\gamma_s$) is defined as the energy needed to separate two contacting solid surfaces per two unit area, which is half of $W_{ss}$. Thus, the magnitude of the surface energy of a solid is an indicator for its cohesiveness. (b) Adhesion between two different materials can make one material (a liquid) spread upon another material (a solid). The contact angle ($\theta$) formed between the two materials is governed by the surface free energies of the two materials ($\gamma_s$ for the solid and $\gamma_l$ for the liquid) and their interfacial tension ($\gamma_{sl}$). The Young-Dupr\'e equation describes the wetting phenomenon between a liquid-solid interface: $\gamma_s=\gamma_{sl}+\gamma_lcos\theta$. Thus, knowing the surface properties of two materials and their interfacial tension can lead to estimation of the wetting scheme/contact angle between them. (c) The surface energy can be partitioned into different components that reflect independent molecular interactions. For example, it can be partitioned into the dispersive component (van der Waals interactions) and the polar component (H-bonding and dipole-dipole interactions). The partitioning of surface energy for a solid material can indicate its bulk chemical make-up.}
\label{fig:sur}
\end{figure}

All of the above phenomena, which occur in Titan's atmosphere and on its surface, are linked to one important physical property of the solid haze particles, surface energy. The surface energy of a solid ($\gamma_s$, unit mJ/m$^2$) is defined as the free energy change when the surface area of a solid is increased a unit area, which is equivalent to the work needed to separate two identical contacting solid surfaces per two unit area, or half of the work of cohesion $W_{ss}$ (Figure \ref{fig:sur}a, Israelachivili, 2011). Thus, knowing the surface energy of the organic haze can provide information about the cohesiveness (or stickiness) of the Titan's aerosol particles and the surface sand grains (assuming the sand is derived from the aerosols and has similar composition). Cohesiveness affects the coagulation and growth of aerosol particles and the mobility of Titan's surface sand, e.g., the threshold wind speed (the minimum wind speed needed to entrain sand particles).

Surface energy also plays an important role in the wetting phenomena. When two dissimilar materials are in contact, the adhesion between these two materials depends on their surface free energies as well as their interfacial energies (Israelachivili, 2011). The wetting behavior of a liquid upon a solid surface can be described by the Young-Dupr\'e Equation (Figure \ref{fig:sur}b, Young, 1805; Dupr\'e, 1869):
\begin{equation}
\gamma_s=\gamma_{sl}+\gamma_lcos\theta,
\label{eq:youngdupre2}
\end{equation}
where $\gamma_s$ and is the surface energy of the solid, $\gamma_l$ is the surface tension of the liquid, $\gamma_{sl}$ is the interfacial tension between the solid and the liquid phase, and $\theta$ is the contact angle or the wetting angle of the liquid on the solid surface. Note that the surface energy of a liquid is often referred as its surface tension, and is denoted with a unit of mN/m (equivalent to mJ/m$^2$). Given the surface energies of the solid and the liquid and the interfacial tension, we could estimate the wetting scheme/contact angle of a liquid on a solid surface. The wetting theory can be applied to many processes happening on Titan. For example, we can use the surface energy of the haze and potential cloud condensates to predict the wetting and nucleation of cloud condensates to understand cloud formation in Titan's atmosphere. We can also predict the contact angle between the solid haze and liquid hydrocarbons to test the viability of a floating organic film on Titan's lakes. 

The surface energy is a macroscopic material property that is determined collectively by various intermolecular interactions between the surfaces (Israelachivili, 2011). Depending on the theory used, the total surface energy can also be divided into different components to reflect specific independent intermolecular forces (Fowkes, 1964). For example, the Owens-Wendt-Rabel-Kaelble (OWRK) model divides the surface energy into dispersive (non-polar) and polar components (Owens \& Wendt, 1969; Rabel, 1971; Kaelble, 1970). The dispersive component includes only London dispersive forces between non-polar molecules, while the polar component includes dipole-dipole and H-bonding interactions. Thus, the magnitude of different components can inform us regarding the chemical nature of the material and allow comparison between different materials (see Figure \ref{fig:sur}c). For example, polyethylene, a non-polar polymer polymerized from C$_2$H$_4$, has a total surface energy of $\sim$33 mJ/m$^2$, with a dispersion component of 33 mJ/m$^2$ and a zero polar component (e.g., Owens \& Wendt, 1969), which means it has minimal polar properties.

\begin{figure}[h]
\centering
\includegraphics[width=33pc]{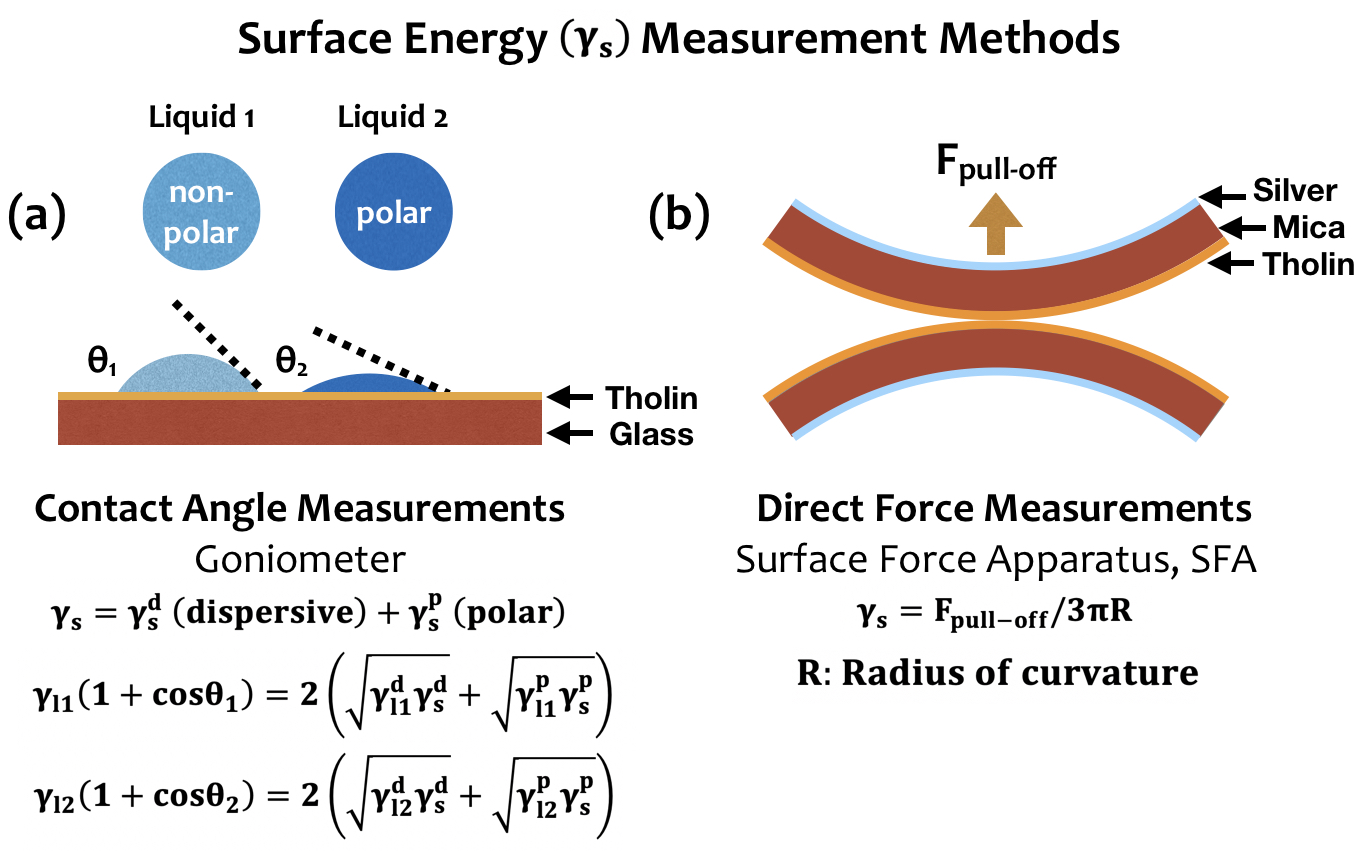}
\caption{Illustrations for surface energy measurement techniques. (a) Sessile drop contact angle method, $\gamma_s$ is the total surface energy of the solid, $\gamma_s^d$ and $\gamma_s^p$ are its dispersive and polar components, $\gamma_{l1}$ and $\gamma_{l2}$ are the total surface tensions of the non-polar and the polar test liquids, and $\gamma_{l1,2}^d$ and $\gamma_{l1,2}^p$ are their dispersive and polar components. (b) Direct force measurement through surface force apparatus. $F_{\rm{pull-off}}$ is the pull-off forces between the two surfaces and R is the radius of curvature of the surfaces.}  
\label{fig:sfa}
\end{figure}

There are multiple ways to determine the surface energy of a material. The sessile drop contact angle measurement retrieves the surface energy of the material indirectly by measuring interactions between the solid material and multiple liquids (see Figure \ref{fig:sfa}(a), e.g., Fowkes, 1964; Zisman, 1964; Owens \& Wendt, 1969; van Oss et al., 1986, 1989; van Oss, 2006). The direct force measurement can give the surface energy of the solid material by measuring the ``pull-off" force between the two materials at contact (see Figure \ref{fig:sfa}(b), e.g., Derjaguin et al., 1975; Johnson et al., 1971; Israelachvili \& McGuiggan, 1990). The surface energy of the Titan aerosol analog, so-called ``tholin", has been measured through the contact angle method with two liquids (Yu et al., 2017a). The total surface energy is measured to be 70.9$_{-4.8}^{+4.6}$ mJ/m$^2$, the dispersion and polar parts are respectively 34.3$_{-2.9}^{+2.7}$ mJ/m$^2$ and 36.6$_{-3.8}^{+3.7}$ mJ/m$^2$. Yu et al. (2017a) used a tholin film produced with cold plasma as the energy source. In this paper, we use additional test liquids which allows for the use of more analytical methods resulting in improved accuracy. Additionally, we use tholin samples produced using a different energy source, a UV lamp, to compare the effect of the energy source on the surface energy of tholin. The UV lamp can simulate solar UV radiation that initiates photochemistry in Titan's atmosphere (Cable et al., 2012). We also used the direct force measurement to measure the total surface energy of cold plasma tholin. The material production and surface preparation is described in Section 2.1. The contact angle and the analytical methods to retrieve the surface energy are described in Section 2.2 and 2.3. The direct force measurement is described in section 2.4. In Section 3.1, the contact angles of test liquids on tholin and their variation with time are summarized and discussed. We present the calculated surface energy for tholins using the contact angle data and compare the results from different analytical methods in Section 3.2. In Section 3.3, we present the measured surface energy of plasma tholin using direct force measurement. Finally, we discuss the applicability of the derived surface energy values from Section 3.2 and 3.3 and their implications for various processes happening on Titan in Section 3.4.

\section{Methods}

\subsection{Material Production}

Tholin samples were produced with two different energy sources, the cold plasma (He et al., 2017) and the UV lamp (He et al., 2018), using the Planetary HAZE Research (PHAZER) experimental system at Johns Hopkins University (apparatus details see He et al., 2017). AC glow discharge is a cold plasma energy source, which is used to simulate the energetic upper atmosphere of Titan. The hydrogen UV lamp (HHeLM-L, Resonance LTD.) produces continuum UV irradiation from 110 and 400 nm with a total UV flux about 3 $\times$ 10$^{15}$ photons/(sr$\cdot$s), which is typically used to simulate photochemistry in Titan's atmosphere (e.g., Trainer et al., 2006, 2012; Sebree et al., 2014; H\"orst et al., 2018).

Prior to the experiments, the substrates were prepared for the contact angle and the direct force measurements. Glass slides (Fisher Scientific) were used as the substrates for the contact angle measurement, they were acid-washed following the procedure described in Yu et al., (2020). For the direct force measurement using the Surface Force Apparatus (SFA), a layer of silver is evaporated onto the backsides of cleaved mica sheets to allow for film thickness, surface deformation, and surface separation measurements by interferometry (Israelachvili \& Adams, 1978). The mica was then glued, silver side down, to the silica support disks using EPON 1004 epoxy. The acid-washed glass slides and the mica surfaces were placed at the bottom of the PHAZER chamber for material deposition.

During material production, a cold gas mixture (5\% $\mathrm{CH_4/N_2}$, 100 K) continuously flows through the reaction chamber with a flow rate of 10 standard cubic centimeters per minute (sccm) and is exposed to the energy source for around 3 s. For the contact angle measurement, the gases flow continuously through the chamber for 72 hr to produce the plasma sample (10 hr for the SFA plasma sample) and 144 hr for the UV sample, which are deposited on the acid-washed glass slides. We ran the UV experiments longer than the cold plasma experiments because the UV experiments usually have much lower yield of tholin (Peng et al., 2013; H\"orst et al., 2018; He et al., 2018) and we need to ensure that we have a homogeneously coated surface of sufficient thickness (over 50 nm) to ensure that contact angle measurement does not depend upon film thickness (e.g., Tavana et al., 2006, Li et al., 2007). The resulting tholin films are relatively smooth for both energy sources as measured by atomic force microscopy (AFM). The root-mean-square (RMS) roughness of the films over an area of 2 $\mu$m $\times$ 2 $\mu$m is measured to be 3--5 nm for the plasma tholin sample and 1--2 nm for the UV tholin sample. For reference, the RMS roughness of the acid-washed bare glass slides is measured to be 0.5--1.5 nm over the 2 $\mu$m $\times$ 2 $\mu$m area. The thickness of the plasma sample is $\sim$1 $\mu$m (Yu et al., 2017a) and is at least 50--100 nm for the UV sample confirmed by AFM. For the direct force measurement, only the plasma sample was made with the gases flowing for 10 hr, leading to a coated film of thickness around 38 nm (measured by interferometry). All samples were stored in a dry N$_2$, oxygen free glove box ($<$0.1 ppm H$_2$O, $<$0.1 ppm O$_2$) to avoid adsorption and surface aging in ambient air prior to the measurements.

\subsection{Sessile drop contact angle measurement}

We used the sessile drop technique to measure the static contact angles between liquid droplets and material surfaces of interest, including uncoated glass slides, cleaved mica sheets, and the coated tholin surfaces. The sessile drop is formed by gently dispensing the test liquids through a pipette onto the material surface in ambient air (temperature 19--20 $^\circ$C, relative humidity 50--60\%). The test liquids used include HPLC-grade water (Fisher Chemical), diiodomethane (ReagentPlus\textsuperscript \textregistered, Sigma-Aldrich, 99\%), glycerol (certified ACS, Fisher Chemical, 99.9\%), ethylene glycol (Aqua Solutions, Inc., 99\%), dimethyl sulfoxide (DMSO, certified ACS, Fisher Chemical, $\geq$99.9\%), formamide (ReagentPlus\textsuperscript \textregistered, Sigma-Aldrich, 99\%), toluene (GR ACS, MilliporeSigma,  $\geq$99.5\%), tetradecane (Sigma-Aldrich, $\geq$99\%), and n-hexane (certified ACS, Fisher Chemical, $\geq$98.5\%). The surface tension values of the test liquids and the corresponding components are adopted from van Oss (2006) and are listed in Table 1.

\begin{table}
\caption{Test liquids and their corresponding surface tensions and surface tension components at 20$^\circ$C (van Oss, 2006), unit is mN/m for all numbers in the table. $\gamma_l$ is the total surface tension. $\gamma^{d}$ and $\gamma^{p}$ are respectively the dispersive and polar components and are used for the OWKR two liquids method and the OWKR fit method. $\gamma^{LW}$, $\gamma^{+}$, and $\gamma^{-}$ are respectively the Liftschitz-van der Waals (LW), electron-acceptor (+), and electron-donor (-) components and are used for the vOCG method.}
\label{table:liquids}
 \begin{center}
 \begin{tabular}{c  c r c  r c c r c c c}
 \toprule
 Liquid & CAS number && Total surface tension$^\ast$ && \multicolumn{2}{c}{OWRK components$^\ast$} && \multicolumn{3}{c}{vOCG components$^\ast$} \\
\cmidrule{4-4} \cmidrule{6-7} \cmidrule{9-11} 
& && $\gamma_l$ && $\gamma_l^d$ & $\gamma_l^p$ && $\gamma_l^{LW}$ & $\gamma_l^+$ & $\gamma_l^-$\\
\midrule
Water &7732-18-5& & 72.8 && 21.8 & 51.0 && 21.8 & 25.5 & 25.5\\
\hline
Glycerol & 56-81-5 & & 64.0 && 34.0 & 30.0 && 34.0 & 3.92 & 57.4\\
\hline
Formamide &75-12-7& & 58.0 && 39.0 & 19.0 && 39.0 & 2.28 & 39.6\\
\hline
Diiodomethane & 75-11-6 & & 50.8 && 50.8 & 0 && 50.8  & 0.01 & 0\\
\hline
Ethylene glycol &107-21-1& & 48.0 && 29.0 & 19.0 && 29.0 & 3.0 & 30.1\\
\hline
DMSO &67-68-5& & 44.0 && 36.0 & 8.0 && 36.0 & 0.5 & 32.0\\
\hline
Toluene &108-88-3& & 28.5 && 28.5 & 0 && 28.5 & 0 & 0.72\\
\hline
Tetradecane &629-59-4& & 26.6 && 26.6 & 0 && 26.6 & 0 & 0\\
\hline
n-Hexane & 110-54-3 & & 18.4 && 18.4 & 0 && 18.4 & 0 & 0\\
\bottomrule
\multicolumn{11}{c}{\footnotesize $^\ast$All units are in mN/m.}
 \end{tabular}
  \end{center}
 \end{table}

\begin{figure}[h]
\centering
\includegraphics[width=32pc]{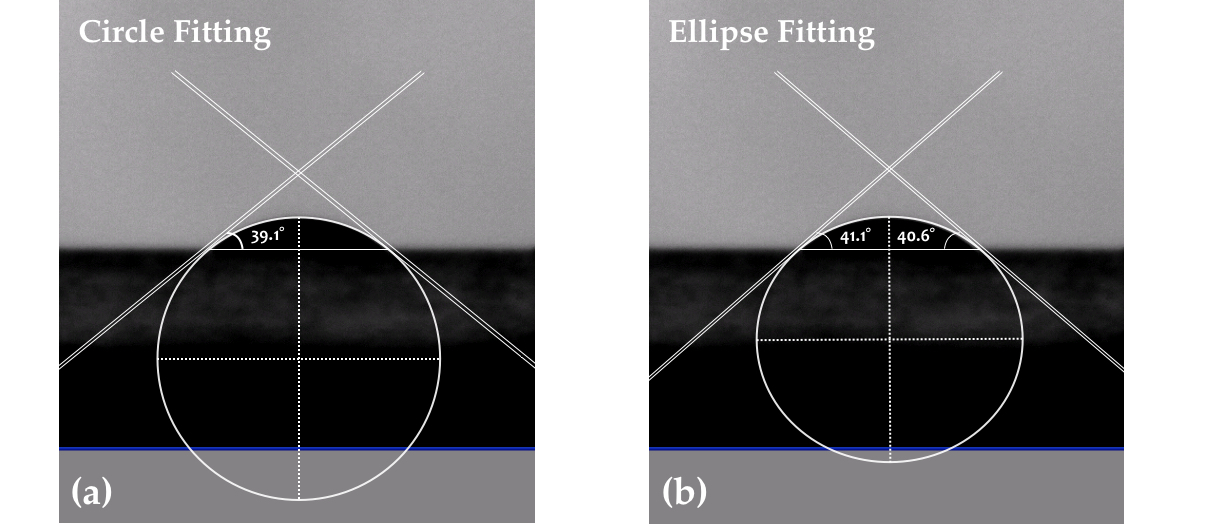}
\caption{(a) Circle fitting for the drop profile formed between diiodomethane and UV tholin. (b) Ellipse fitting for the same drop profile formed between diiodomethane and UV tholin.}
\label{fig:circle}
\end{figure}

We performed the measurements for each test liquid on the material surfaces two to six times on different areas of the film. The drop size is controlled to be 2 $\mu$L to avoid the flattening effect due to gravity for large droplets (Extrand \& Moon, 2010; Zhang et al., 2008; McGuiggan et al., 2011). We used a Ram\'e-Hart goniometer to capture focused images of the droplet with a sharp boundary and a clear liquid-solid interface contact line from the side. The static contact angle was measured on the image by using the ImageJ software (Schneider et al., 2012) with the contact angle plugin. We manually chose two points to define the baseline and five points to define the drop profile. The contact angle plugin fits the profile of the drop with either circle or ellipse approximation. The circle approximation is commonly used for small droplets or droplets with small contact angle. The ellipse approximation can more realistically fit larger droplets and droplets with larger contact angles, where the flattening of the drop profile due to gravity is more pronounced (Xu, 2012). Examples of circle and ellipse fitting for the same droplet profile between diiodomethane and UV tholin are shown in Figure \ref{fig:circle}(a) and (b). To validate our methodology, we measured the contact angles between selected test liquids (diiodomethane and water) and two well-known surfaces, cleaved mica sheets, which is smooth on the molecular scale, and the uncoated glass slides, the results are shown in Table \ref{table:contactangle_sample}. The results are compatible to previous contact angle measurements (e.g., Shafrin \& Zisman, 1967; Wu, 1982; Liberelle et al., 2008).

\begin{table}
\caption{Contact angle between selected test liquids and two well-known surfaces, the substrate glass slide and cleaved mica sheets. All numbers have units of degree. The contact angle was measured through both circle fitting and ellipse fitting. The standard deviation (SD) for each fitting is also recorded.}
 \label{table:contactangle_sample}
 \centering
 \begin{tabular}{ c c c c c r cccc }\toprule
Liquid & \multicolumn{4}{c}{Glass slide}  && \multicolumn{4}{c}{Cleaved mica}\\
 \cmidrule{2-5}  \cmidrule{7-10}
& Circle fitting & SD & Ellipse fitting & SD && Circle fitting & SD & Ellipse fitting & SD \\
\midrule
Water & 13.7$^\circ$ & 3.2$^\circ$ & 15.4$^\circ$ & 3.4$^\circ$  && spread & n/a & spread  & n/a  \\
Diiodomethane & 36.1$^\circ$ & 1.1$^\circ$ & 39.0$^\circ$ & 2.0$^\circ$  &&  38.8$^\circ$ & 1.5$^\circ$ & 39.4$^\circ$ & 1.7$^\circ$ \\
\bottomrule
 \end{tabular}
 \end{table}

\subsection{Surface energy derivation methods--contact angle measurement}
Equation \ref{eq:youngdupre2}, the Young-Dupr\'e equation (Young, 1805; Dupre, 1869) describes the mechanical energy balance between the solid and the liquid phases at the contact line. In Equation \ref{eq:youngdupre2}, The interfacial tension ($\gamma_{sl}$) can also be written as the sum of the surface free energies minus the work of adhesion ($W_{sl}$) between the two phases, $\gamma_{sl}=\gamma_s+\gamma_l-W_{sl}$, thus the Young-Dupr\'e equation can also be written as:
\begin{equation}
W_{sl}=\gamma_l(1+cos\theta),
\label{eq:youngdupre}
\end{equation}
where $W_{sl}$ is the work of adhesion between the liquid and the solid, which describes the energy needed to separate the solid and liquid. In Equation \ref{eq:youngdupre}, only $\gamma_l$ and $\theta$ are known or directly measurable quantities, while the solid surface energy is included in the work of adhesion term $W_{sl}$. In order to use contact angle measurement to understand the origins of the surface energy, the surface tension component method was proposed (Fowkes, 1964). The theory assumes that: 1) the total surface free energy can be partitioned into different individual components representing contributions from different intermolecular forces; 2) the work of adhesion $W_{sl}$ can be expressed as the geometric mean of the surface tension components. The Fowkes method only considers dispersion components and thus is not applicable to polar materials. The Owens-Wendt-Rabel-Kaelble (OWRK) method includes both the dispersion components and the polar components (including dipole-dipole and H-bonding interactions) and is currently the most widely used surface energy derivation method for contact angle measurement (OWRK two-liquid method). Here we also included 1) an extended version of OWRK method to incorporate contact angle data from more liquids (OWRK-fit method), 2) the Wu method which partitions the work of adhesion into the harmonic mean of the surface energy components (Wu, 1971), and 3) the van Oss, Chaudhury, and Good (vOCG) method which partitions the total surface free energy into the dispersive, electron-acceptor, and electron-donor components (van Oss et al., 1986, 1989; van Oss, 2006). Finally, an empirical approach that correlates the cosine of the contact angle and the total surface tension of the liquids, called the Zisman method (Zisman, 1964), is used to derive the critical surface tension ($\gamma_c$) of the solid.

The OWRK method partitions the total surface free energy of the solid and the liquid into the dispersion and the polar components:
\begin{align}
\begin{aligned}
\gamma_s=\gamma_s^d+\gamma_s^p, \\
\gamma_l=\gamma_l^d+\gamma_l^p,
\label{eq:owrk1}
\end{aligned}
\end{align}
where $\gamma_s$ is the total surface energy of the solid, $\gamma^d_s$ and $\gamma^p_s$ are respectively the dispersion and the polar components of the surface energy of the solid, and $\gamma^d_l$ and $\gamma^p_l$ are respectively the dispersion and the polar component of the surface tension of the test liquid. The work of adhesion $W_{sl}$ can be approximated as the geometric means of the surface tension components:
\begin{equation}
W_{sl}=2(\sqrt{\gamma^d_s\gamma^d_l}+\sqrt{\gamma^p_s\gamma^p_l}).
\label{eq:finalwork}
\end{equation}

With Equation \ref{eq:youngdupre} and \ref{eq:finalwork}, we have:
\begin{equation}
\gamma_{li}(1+cos\theta_{i})=2(\sqrt{\gamma^d_s\gamma^d_{li}}+\sqrt{\gamma^p_s\gamma^p_{li}}),
\label{eq:contactangle}
\end{equation}
where the subscript i refers to a specific liquid. After we measured the contact angles of the solid with two test liquids, we can solve two sets of equation \ref{eq:contactangle} to get the dispersive and the polar part of the solid surface energy. The result of this method is dependent upon the choice of liquids, and the combination of water and diiodomethane in general provides the most accurate results (Hejda et al., 2010).  

When there are contact angle measurements for more than two test liquids, we can convert Equation \ref{eq:contactangle} to the following by dividing both sides of the equation by 2$\sqrt{\gamma^d_{li}}$:
\begin{equation}
\frac{\gamma_{li}(1+cos\theta_{i})}{2\sqrt{\gamma^d_{li}}}=\sqrt{\gamma^d_s}+\sqrt{\frac{\gamma^p_{li}}{\gamma^d_{li}}}\sqrt{\gamma^p_s},
\label{eq:contactangle2}
\end{equation}
so we can incorporate the data for more than two test liquids into a linear fit (Hejda et al., 2010), which is also called an OWRK-fit method. The left side of the equation becomes $y_i$ and $\sqrt{\gamma^p_{li}/\gamma^d_{li}}$ becomes $x_i$. The square root of the polar and dispersion components of the solid surface energy can then be derived as the linear regression coefficients.

The Wu method (Wu, 1971) uses the harmonic mean instead of the geometric mean to partition the work of adhesion $W_{sl}$:
\begin{equation}
W_{sl}=4(\frac{\gamma^d_{li}\gamma^d_s}{\gamma^d_{li}+\gamma^d_s}+\frac{\gamma^p_{li}\gamma^p_s}{\gamma^p_{li}+\gamma^p_s}).
\label{eq:Wu}
\end{equation}
With Equation \ref{eq:youngdupre} and \ref{eq:Wu}, we have:
\begin{equation}
\gamma_{li}(1+cos\theta_{i})=4(\frac{\gamma^d_{li}\gamma^d_s}{\gamma^d_{li}+\gamma^d_s}+\frac{\gamma^p_{li}\gamma^p_s}{\gamma^p_{li}+\gamma^p_s}).
\label{eq:Wumethod}
\end{equation}
The Wu method needs two sets of contact angle data from two liquids to solve Equation \ref{eq:Wumethod} and the result is also dependent on the choice of liquids.

The vOCG method (van Oss et al., 1986, 1989; van Oss, 2006) partitions the surface free energy into three components, the dispersion or the Lifshitz-van der Waals (LW) component which includes the London dispersion force, Keesom dipole-dipole force, and the Debye dipole-induced dipole force, the electron-acceptor component, and the electron-donor component:
\begin{equation}
\gamma_{s}=\gamma^{LW}_{s}+\gamma^{AB}_{s}=\gamma^{LW}_{s}+2\sqrt{\gamma^{+}_{s}\gamma^{-}_{s}},
\label{eq:voCG2}
\end{equation}
where $\gamma^{LW}$ is the Lifshitz-van der Waals (LW) component (or the dispersive component since London dispersion force dominates), $\gamma^{AB}$ is the Lewis acid-base component (or the polar component), which can be expressed by an asymmetrical combination of the electron-acceptor ($\gamma^{+}$) and the electron-donor ($\gamma^{-}$) components. The work of adhesion is expressed as:
\begin{equation}
W_{sl}=2(\sqrt{\gamma^{LW}_{li}\gamma^{LW}_{s}}+\sqrt{\gamma^{+}_{li}\gamma^{-}_{s}}+\sqrt{\gamma^{-}_{li}\gamma^{+}_{s}}).
\label{eq:voCG1}
\end{equation}
With Equation \ref{eq:youngdupre} and \ref{eq:voCG1}, we have:
\begin{equation}
\gamma_{li}(1+cos\theta_{i})=2(\sqrt{\gamma^{LW}_{li}\gamma^{LW}_{s}}+\sqrt{\gamma^{+}_{li}\gamma^{-}_{s}}+\sqrt{\gamma^{-}_{li}\gamma^{+}_{s}}).
\label{eq:voCG}
\end{equation}
Equation \ref{eq:voCG} has three unknowns and thus needs contact angle data from three test liquids to solve the solid surface energy.

Fox and Zisman (1950) found an empirical linear relationship between the cosine of the contact angle (cos$\theta$) and the surface tension of the liquid ($\gamma_l$) for certain polymers. By extrapolating the linear function of cos$\theta$ and $\gamma_l$ to cos$\theta=1$, a critical surface tension of the solid ($\gamma_c$) can be found. It is the highest surface tension of a liquid that would completely wet the solid surface, and is empirically very close to the surface energy of the solid ($\gamma_s$).

The OWRK two-liquid, Wu, and voCG methods are essentially very similar in that they all partition the surface tension into different components, and the surface energies of the solid can be derived using the contact angle results of two or three liquids. However, these methods are all sensitive to the choice of liquids, for example, Shimizu \& Demarquette (1999) found that different choice of liquid could lead to 60--130\% variation in the derived surface energy values of a solid. However, not all liquid pairs would yield meaningful results. Using a pair of non-polar liquids would lead to a zero polar component in the surface energy calculation, as non-polar liquids cannot probe the polar properties of the solid. Using a liquid with too low of a surface tension also may not yield useful results as low surface tension liquids would readily wet a solid surface with high surface energy, and thus cannot fully probe the surface energy of the solid. So, for the two liquid methods (OWRK two-liquid and Wu method), we choose diiodomethane and water as the test liquid pair, as among the solvents used in this study, diiodomethane has the highest dispersive component (50.8 mN/m), and water has the highest polar component (51.0 mN/m), so this pair can fully probe both the dispersive and the polar properties of the solid surface. For the voCG method, we choose water, diiodomethane, and one other polar liquid (DMSO, ethylene glycol, formamide, or glycerol) as the test liquids.

The OWRK-fit and the Zisman plot methods are able to incorporate the contact angle results for all test liquids. However, the OWRK-fit may not be able to fully probe the polar/dispersive properties of a high surface energy solid when most of test liquids have lower surface tensions compared to the solid. The Zisman plot method is limited to mostly low surface energy surfaces as the linear relationship is only attributed to dispersive interactions (Zisman, 1964) and may break if the solid has significant polar properties.

\subsection{Direct force measurement--surface force apparatus}

In addition to the contact angle measurement, the SFA was used to directly measure the adhesion force between mica surfaces coated with cold plasma tholin, which was then used to derive the surface energy of tholin based on its definition. The SFA has the advantage over the AFM experiments used in Yu et al., (2017a) since the distance between the surfaces and the contact area can be directly measured at the area of interest which is near and within the contact zone. The plasma tholin coated mica surfaces prepared in Section 2.1 were mounted in the SFA and a small vial of P$_2$O$_5$ was placed in the chamber. The P$_2$O$_5$ is used to absorb water from the atmosphere. The chamber was quickly closed and dry nitrogen was purged through the chamber for 15 hours.

\begin{figure}[h]
\centering
\includegraphics[width=33pc]{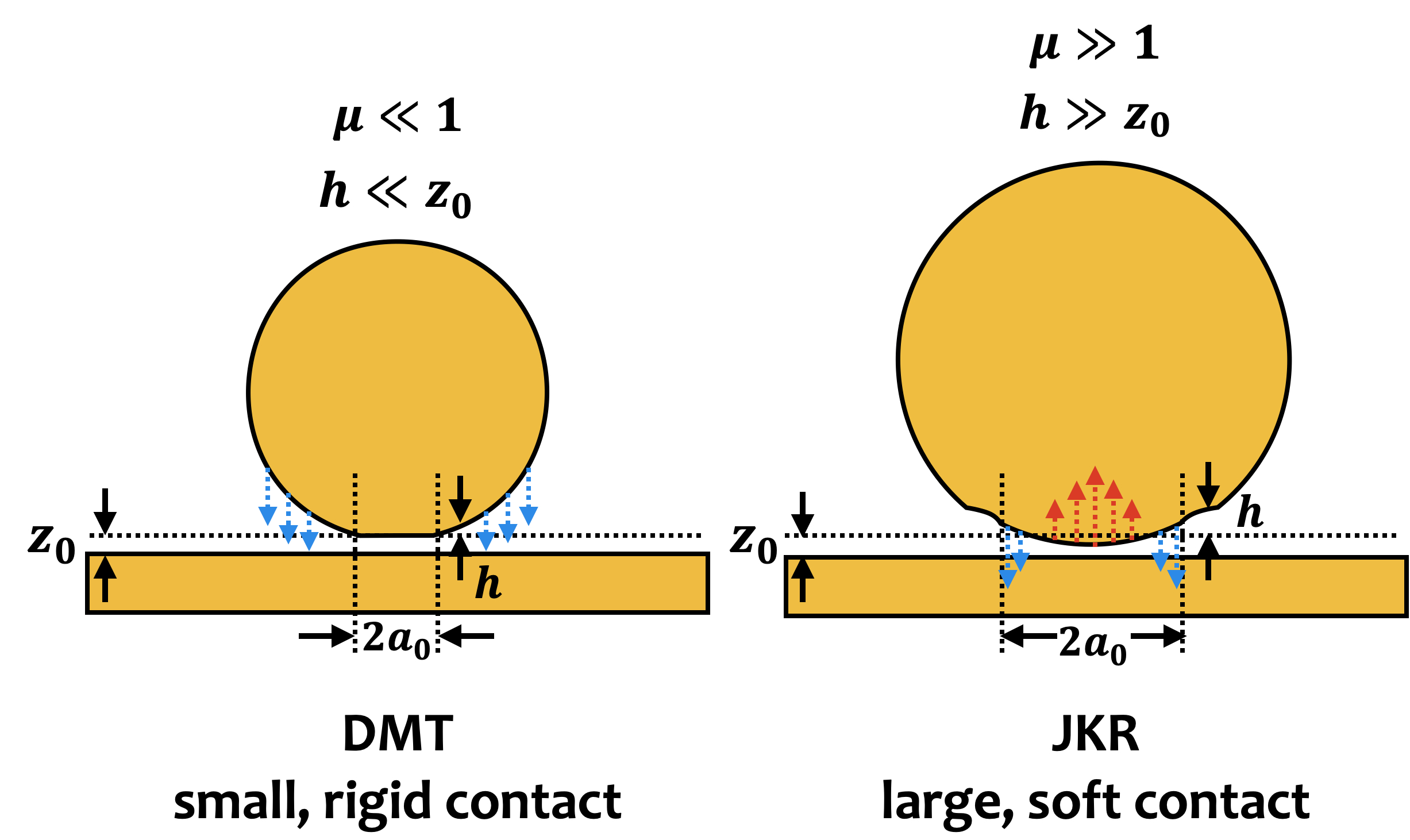}
\caption{Exaggerated contact schemes between a sphere and a flat surface for the DMT and the JKR theories when the load is zero, and the Tabor parameter ($\mu$) used to determine the applicability of the two contact theories. The contact zone has a radius of $a_0$. The gap outside the contact zone has a height of $h$. The dashed arrows mark the net forces at different points of the contact. The net force equals to zero at the equilibrium atomic separation distance ($z_0$). The net force is repulsive (red dashed arrow pointing upward) when the separation distance between the contact points is smaller than $z_0$, and is attractive (blue dashed arrow pointing downward) when the separation distance between the contact points is larger than $z_0$. For the DMT contact scheme, only long-range forces outside the contact zone are accounted for, where $h$ is smaller or comparable to $z_0$ ($\mu\ll1$). For the JKR contact scheme, forces outside the contact zone are neglected as $h\gg z_0$ ($\mu\gg1$), and only short-range forces inside the contact zone are accounted for.}
\label{fig:contact}
\end{figure}

During the SFA measurements, the two coated tholin surfaces were first brought together. An adhesion force was measured by separating the surfaces. During the separation, the contact diameter was continuously monitored by interferometry until the surfaces jump out of contact, where a ``pull-off" force can be measured (F${\rm{_{pull-off}}}$). The relationship between the pull-off force and the surface energy can be described by two theories, the Johnson-Kendall-Roberts (JKR) theory (Johnson et al., 1971) for large, soft contacts and the Derjaguin, Muller, and Toporov (DMT) theory (Derjaguin et al., 1975) for small, hard contacts. The Tabor parameter is used to determine the contact mechanics regime (Figure \ref{fig:contact}). It is defined as the ratio between the height of the gap ($h$) outside the contact zone and the equilibrium separation distance between atoms ($z_0$), where h is given by (Tabor, 1977):
\begin{equation}
h\simeq(\frac{RW^2}{E^2_*})^{1/3},
\label{eq:tabor2}
\end{equation}
where R is the radius of curvature of the surfaces, W is the work of adhesion between the two surfaces, for two identical surfaces, $W=2\gamma_{sv}$, $E_*$ is the effective elastic modulus. The net van der Waals force is zero at $z_0$. When h is smaller or comparable with $z_0$, interactions outside the contact zone need to be accounted for (DMT theory), while when $h$ is much larger than $z_0$, interactions within the contact zone dominate (JKR theory). The Tabor parameter is thus defined as a dimensionless number $\mu$ (Tabor, 1977):
\begin{equation}
\mu=\frac{h}{z_0}=(\frac{4R\gamma^2_{sv}}{E^{2}_*z^3_0})^{1/3}.
\label{eq:tabor}
\end{equation}
The effective modulus $E_*$ for this system is around 30 GPa, and $z_0$ is is around 0.3 nm (McGuiggan et al., 2007).

Typically when the Tabor parameter is greater than 5, the JKR theory is applicable. Because the radii of the mica surfaces are relatively large (R = 1.3 cm), the Tabor parameter is very large and the JKR theory applies. Using the JKR approximation, the pull-off force can be used to compute the surface energy ($\gamma_{sv}$) of the adhesive surfaces:
\begin{equation}
F_{\rm{pull-off}}=3\pi R \gamma_{sv}.
\label{eq:jkr}
\end{equation}

\section{Results}

\subsection{Contact angle between tholin and test liquids}

Table \ref{table:contactangle} summarizes the measured contact angles between test liquids and the tholin samples. Liquids with the lowest surface tensions, toluene, tetradecane, and n-hexane, completely spread on both tholin samples with a zero contact angle. The measured contact angle from both circle and ellipse fittings are presented in Table \ref{table:contactangle}. Both fits yield similar results within the standard deviations of each other. In a similar contact angle study (Lamour et al., 2010), circle fitting was used for droplets with contact angles less than 40$^\circ$. The measured contact angles in our study are mostly within this range, so we will use the contact angle obtained from circle fitting for calculations in the following sections.

\begin{table}
\caption{Contact angles between the test liquids and the tholin samples, measured immediately after the droplet formed on the sample (t = 0 s), all numbers have units of degree. The contact angle was measured through both circle fitting and ellipse fitting. The standard deviation (SD) for each fitting is also recorded.}
 \label{table:contactangle}
 \centering
 \begin{tabular}{ c c c c c r cccc }\toprule
Liquid & \multicolumn{4}{c}{Plasma tholin}  && \multicolumn{4}{c}{UV tholin}\\
 \cmidrule{2-5}  \cmidrule{7-10}
& Circle fitting & SD & Ellipse fitting & SD && Circle fitting & SD & Ellipse fitting & SD \\
\midrule
Water & 31.2$^\circ$ & 3.6$^\circ$ & 33.2$^\circ$ & 5.0$^\circ$  && 36.7$^\circ$  & 8.3$^\circ$  & 37.5$^\circ$  & 7.1$^\circ$ \\
Glycerol & 42.3$^\circ$ & 5.3$^\circ$ & 41.6$^\circ$ & 1.1$^\circ$  && 53.1$^\circ$ & 0.7$^\circ$  & 51.2$^\circ$ & 3.0$^\circ$ \\
Formamide & 28.2$^\circ$  & 8.3$^\circ$ & 29.0$^\circ$ & 6.0$^\circ$  && 40.5$^\circ$ & 0.7$^\circ$  & 41.6$^\circ$ & 1.3$^\circ$\\
Diiodomethane & 40.0$^\circ$ & 4.0$^\circ$ & 43.5$^\circ$ & 3.0$^\circ$  &&  37.0$^\circ$ & 2.5$^\circ$ & 38.2$^\circ$ & 2.9$^\circ$ \\
Ethylene glycol & 35.3$^\circ$  & 5.4$^\circ$  & 36.8$^\circ$  & 3.7$^\circ$  && 26.1$^\circ$ & 1.4$^\circ$ &  28.3$^\circ$ & 2.4$^\circ$\\
DMSO & 27.3$^\circ$  & 7.5$^\circ$ & 27.1$^\circ$ & 6.3$^\circ$  && 16.3$^\circ$ & 9.2$^\circ$  & 19.3$^\circ$ & 4.3$^\circ$ \\
Toluene & spread & n/a & spread & n/a && spread &  n/a & spread & n/a \\
Tetradecane & spread & n/a &  spread  & n/a  && spread & n/a  & spread & n/a\\
n-Hexane & spread & n/a &  spread  & n/a  && spread & n/a  & spread & n/a\\
\bottomrule
 \end{tabular}
 \end{table}

\begin{figure}[h]
\centering
\includegraphics[width=33pc]{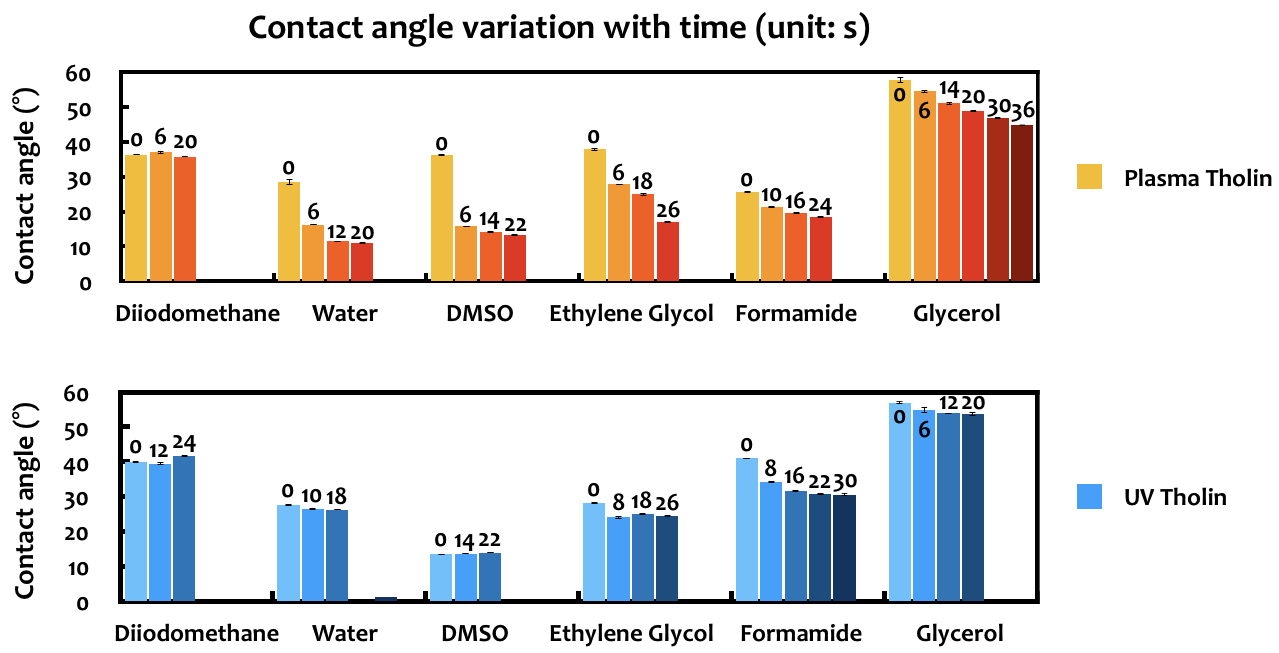}
\caption{The variation of the measured contact angle with time for different liquids. The top histogram is for plasma tholin and the bottom histogram is for UV tholin. All the marked times are in seconds.}
\label{fig:time}
\end{figure}

Figure \ref{fig:time} plots the change in contact angle versus time for different liquids tested on the tholin samples. The equilibrium timescale for the liquid drop to reach the static state is usually less than a second (Law and Zhao, 2016), thus the images of the drop profile were captured as soon as the liquids wet the solid surface (results shown as the contact angle marked at 0 s). The contact angles between the non-polar diiodomethane and both tholin samples are stable with time. While for plasma tholin, the contact angle with all the polar liquids changes drastically with time, indicating that it is soluble in these polar solvents, which has been observed in other plasma samples (McKay, 1996; Coll et al., 1999; Sarker et al., 2003; Carrasco et al., 2009; Kawai et al., 2013; He \& Smith, 2014a). The solubilities of a plasma tholin sample produced under similar conditions (same initial gas mixture and energy source) as our plasma sample was measured previously by He \& Smith (2014a), which gives a 3--4 mg/mL of solubility in water and a 126.2 mg/mL of solubility in DMSO. The contact angles between UV tholin and polar solvents including water and DMSO are stable with time, which corresponds to its insolubility in these two solvents (which has been observed in other UV samples, e.g., Joseph, 1999).

\subsection{The surface energy of tholin--contact angle measurement}

The surface energy of the plasma and the UV tholins are calculated with multiple analytical methods as described in Section 2.3, the results are summarized in Table \ref{table:surfaceenergy}. Even though the contact angle results are different for plasma tholin and UV tholin, as shown in Table \ref{table:contactangle}, the calculated surface energies for both tholin are similar for the same method (within 10 mJ/m$^2$ considering the standard deviations). It is interesting that the total surface energy values for the tholins made with different energy sources (plasma and UV) are relatively close to each other even though their solubilities are very different. The plasma and UV tholins have similar dispersive components, and UV tholin has lower polar components compared to plasma tholin. This is likely caused by the differences in the energy sources, as cold plasma can directly break both nitrogen and methane, while the UV lamp used in our experiments (110--400 nm) cannot directly dissociate the triple-bonded nitrogen in the gas mixture (He et al., 2019), and nitrogen is likely incorporated in the solid sample through other secondary processes (Trainer et al., 2012). Tholins also have higher surface energies compared to common polymers, whose surface energies are usually between 20--50 mJ/m$^2$ (Owens \& Wendt, 1969; Wu, 1971; Wu, 1982; van Oss, 2006). The surface energies of tholins are closer to the surface energy of amorphous carbon (59$\pm$3 mJ/m$^2$), which is measured by the same contact angle method and derived with the OWRK two-liquid method (McGuiggan et al., 2002). 

\begin{figure}[h]
\centering
\includegraphics[width=28pc]{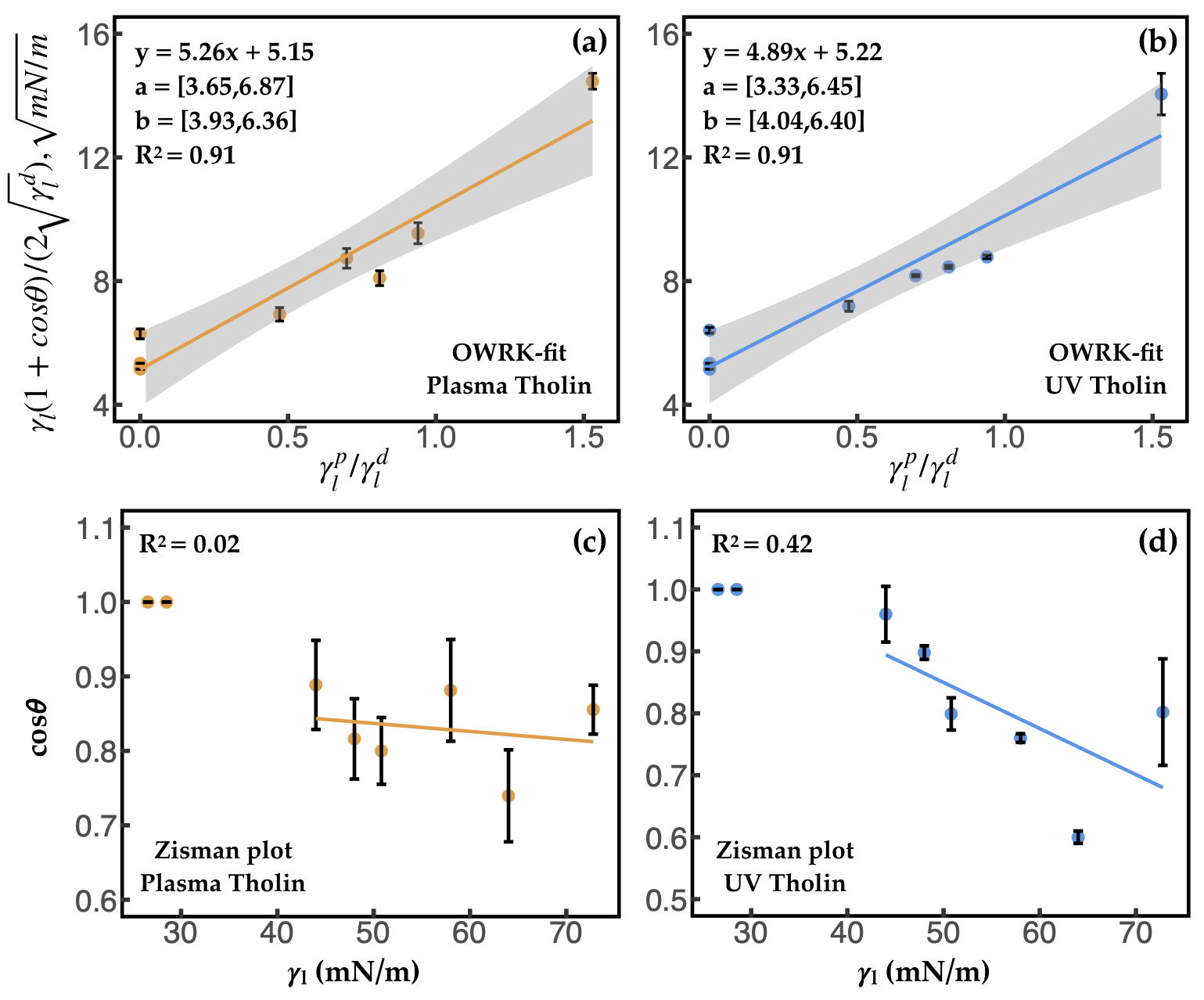}
\caption{(a) OWRK-fit for plasma tholin. The data points x$_i$s are calculated from Table \ref{table:liquids} for different test liquids (from left to right: (diiodomethane, toluene, tetradecane), DMSO, formamide, ethylene glycol, glycerol, water), and y$_i$s are calculated with the contact angle data from Table \ref{table:contactangle} using Equation \ref{eq:contactangle2}. The orange line shows the linear fit to the data (y = ax + b). R$^2$ is the coefficient of determination of the linear fit. The slope, a, gives the square root of the polar component of the solid surface energy and the intercept, b, gives the square root of the dispersive component. The gray shaded area marks the 95\% confidence intervals for the fitting. The range for the slope, a, and the intercept, b, within the 95\% confidence are given as well. (b) OWRK-fit for UV tholin, where the blue line shows the linear fit to the data. (c) Zisman plot for plasma tholin, x$_i$ is the total surface tension of the test liquids (from left to right: tetradecane, toluene, DMSO, ethylene glycol, diiodomethane, formamide, glycerol, water), and y$_i$ is the contact angle between the plasma tholin sample and each test liquid. The orange line gives the linear fit to the data. (d) Zisman plot for UV tholin, where the blue line shows the linear fit to the data.}
\label{fig:zisman_ow}
\end{figure}

We use five different analytical methods to derive the surface energy of tholin. The OWRK two-liquid and the Wu methods give the highest surface energy values, around 65--70 mJ/m$^2$. The vOCG method gives the lowest value (as low as $\sim$40 mJ/m$^2$ for UV tholin). OWRK-fit method gives intermediate value of around 50--55 mJ/m$^2$ (linear fits are shown in panel (a) and (b) of Figure \ref{fig:zisman_ow}). The Zisman plot method does not seem to yield a reasonable result for both tholin samples, as the coefficients of determination (R$^2$) for both linear fits are less than 0.5 (see panel (c) and (d) of Figure \ref{fig:zisman_ow}).

The OWRK two-liquid and Wu methods give similar results using the same pair of test liquids (diiodomethane and water), as they both partition the surface energy into two components (dispersive and polar, see Equation \ref{eq:owrk1}. The difference is that OWRK two-liquid method partitions the work of adhesion using the geometric mean of the surface energy components (Equation \ref{eq:finalwork}), while the Wu method uses the harmonic mean (Equation \ref{eq:Wu}). In general, the surface energy of a solid derived using these two methods deviates at most a few mJ/m$^2$ from each other (e.g., Shimuzu \& Demarquette, 2000). The total tholin surface energies derived using the OWRK two-liquid method ($\gamma_s=$ 66--68 mJ/m$^2$) are relatively high compared to most polymers derived using the same method (20--50 mJ/m$^2$, e.g., Owens \& Wendt, 1969), and tholins also have higher surface energies derived from the Wu method ($\gamma_s=$ 70--72 mJ/m$^2$) compared to most polymers derived with the same method (20--50 mJ/m$^2$, e.g., Wu, 1971, Wu, 1982). The high surface energy measured for tholin is expected as large fraction of polar species were found in tholin (e.g., Quirico et al., 2008; Derenne et al., 2012; He et al., 2012; He \& Smith, 2014b; Maillard et al., 2020).

The voCG method is different from the Wu and OWRK two-liquid method in that it partitions the surface energy into three components (Lifshitz-van der Waals, electron-donor and electron-acceptor) and two components are non-additive (see Equation \ref{eq:voCG2}). Using this method, van Oss (2006) found that most polymers have the polar part of their surface energy dominated by electron-donor components ($\gamma^-_s$) with minimal electron-acceptor components ($\gamma^+_s\sim0$). Both plasma and UV tholins are strong electron-donors, with plasma tholin having some electron-acceptor properties ($\gamma^+_s=$ 0.6--0.8 mJ/m$^2$). Tholins are stronger electron-donors ($\gamma^-_s=$ 45--50 mJ/m$^2$) compared to most polymers derived from voCG method ($\gamma^-_s<$ 40 mJ/m$^2$, van Oss, 2006). The voCG method, however, does not yield results in positive real numbers for some of the liquids triplets tested, including diiodomethane-ethylene glycol-water for both tholins and diidomethane-DMSO-water for plasma tholin.

The OWRK-fit method uses the same principle as the OWRK two-liquid method (Equation \ref{eq:owrk1}), but is able to incorporate contact angle data for more test liquids. It yields similar polar component values ($\gamma^{p}_s=$ 27.7 mJ/m$^2$ for plasma tholin and $\gamma^{p}_s=$ 23.9 mJ/m$^2$ for UV tholin) compared to the OWRK two-liquid method ($\gamma^{p}_s=$ 28.5 mJ/m$^2$ for plasma tholin and $\gamma^{p}_s=$ 24.9 mJ/m$^2$ for UV tholin), but lower dispersive component values ($\gamma^{p}_s=$ 26.5 mJ/m$^2$ for plasma tholin and $\gamma^{p}_s=$ 27.3 mJ/m$^2$ for UV tholin) than the two-liquid method ($\gamma^{p}_s=$ 39.6 mJ/m$^2$ for plasma tholin and $\gamma^{p}_s=$ 41.1 mJ/m$^2$ for UV tholin). But the 95\% confidence interval for the linear fit gives a much wider range of surface energy for the solid, the dispersive component of plasma tholin ranges from 15.4 to 40.4 mJ/m$^2$, and its polar component ranges from 13.3 to 47.2 mJ/m$^2$; for UV tholin, its dispersive component ranges from 16.3 to 41.0 mJ/m$^2$, and its polar component ranges from 11.1 to 41.6 mJ/m$^2$ (see Figure \ref{fig:zisman_ow}). Note that diiodomethane and water are the leftmost and rightmost points on the OWRK-fit plot, and a line that connects these two farthest points would establish the general trend of the linear fit (which is equivalent to the OWRK two-liquid method). This proves again that using diiodomethane and water as the two-liquid pair would lead to the most significant results for the two-liquid methods (Hedja et al., 2010). The polar component from the slope of the OWRK-fit (see Figure \ref{fig:zisman_ow}(a) and (b)) is similar to the slope established between diiodomethane and water, and the dispersive component from the intercept of the linear fit is lower compared to the intercept for the line between diiodomethane and water. The OWRK-fit method is able to incorporates data for all test liquids but may suffer from its inability to fully probe the polar/dispersive properties of high surface energy solid when low surface tension liquids are used, such as toluene and tetradecane, which both completely wet both tholin surfaces. 

The Zisman plot method did not yield meaningful results for both plasma and UV tholin since there is no good linear fit for the contact angle data. This tells us that tholin is a polar solid with high surface energy, as this method generally applies for low surface energy solids where van der Waals interaction dominates (e.g. Hedja et al., 2010, Law \& Zhao, 2016).

\begin{table}
\caption{The derived surface energy of the tholin samples and the different components. All numbers have units of mJ/m$^2$. The standard deviations (SD) for OWRK two-liquid method and the Wu method are calculated through propagation of error derived from Equation \ref{eq:contactangle} and \ref{eq:Wumethod}.}
 \label{table:surfaceenergy}
 \centering
 \begin{tabular}{ c c c c c cc r cccc cc}\toprule
 Methods & \multicolumn{13}{c}{Derived total surface energy and surface energy components$^\ast$}  \\
  \cmidrule{2-14} 
 & \multicolumn{6}{c}{Plasma tholin}  && \multicolumn{6}{c}{UV tholin}\\
 \cmidrule{2-7}  \cmidrule{9-14}
& $\gamma_s$ & SD &  $\gamma^d_s$ & SD  & $\gamma^p_s$ & SD && $\gamma_s$ & SD & $\gamma^d_s$ & SD & $\gamma^p_s$ & SD \\
\midrule
Yu et al., (2017a) &  \multirow{2}{*}{70.9} &  \multirow{2}{*}{4.8} & \multirow{2}{*}{34.3} &  \multirow{2}{*}{2.8} &  \multirow{2}{*}{36.6} &  \multirow{2}{*}{3.8}   &&  \multirow{2}{*}{n/a}  &  \multirow{2}{*}{n/a}  &  \multirow{2}{*}{n/a}  &  \multirow{2}{*}{n/a} &  \multirow{2}{*}{n/a} &  \multirow{2}{*}{n/a} \\
OWRK (diiodomethane-water)&  &  &  &   &&   &  &  &  \\
\hline
OWRK two-liquid method &  \multirow{2}{*}{68.1} &  \multirow{2}{*}{2.5} & \multirow{2}{*}{39.6} &  \multirow{2}{*}{1.4} &  \multirow{2}{*}{28.5} &  \multirow{2}{*}{2.2}   &&  \multirow{2}{*}{66.0} &  \multirow{2}{*}{4.7} & \multirow{2}{*}{41.1}  &  \multirow{2}{*}{0.9}  &  \multirow{2}{*}{24.9}  &  \multirow{2}{*}{4.7}  \\
(diiodomethane-water)&  &  &  &   &&   &  &  &  \\
\hline
Wu method &  \multirow{2}{*}{72.1} &  \multirow{2}{*}{2.5} & \multirow{2}{*}{40.2} &  \multirow{2}{*}{1.8} &  \multirow{2}{*}{31.9} &  \multirow{2}{*}{1.7}   && \multirow{2}{*}{70.5} &  \multirow{2}{*}{4.8} & \multirow{2}{*}{41.5}  &  \multirow{2}{*}{1.1}  &  \multirow{2}{*}{29.0}  &  \multirow{2}{*}{3.9} \\
(diiodomethane-water)&  &  &  &   &&   &  &  &  \\
\hline
OWRK-fit &  \multirow{2}{*}{54.1} &  \multirow{2}{*}{} & \multirow{2}{*}{26.5} &  \multirow{2}{*}{} &  \multirow{2}{*}{27.7} &  \multirow{2}{*}{}   && \multirow{2}{*}{51.2} &  \multirow{2}{*}{} & \multirow{2}{*}{27.3}  &  \multirow{2}{*}{}  &  \multirow{2}{*}{23.9}  &  \multirow{2}{*}{}  \\
(all test liquids) &  &  &  &   &&   &  &  &  \\
\midrule
& $\gamma_s$  & &  $\gamma^{LW}_s$ & $\gamma^{+}_s$  & $\gamma^{-}_s$ &  $\gamma^{AB}_s$  &&  $\gamma_s$ & & $\gamma^{LW}_s$ & $\gamma^{+}_s$ & $\gamma^{-}_s$ & $\gamma^{AB}_s$ \\
\midrule
vOCG method &  \multirow{2}{*}{50.5} &  \multirow{2}{*}{} & \multirow{2}{*}{38.4} &  \multirow{2}{*}{0.8} &  \multirow{2}{*}{45.5} &  \multirow{2}{*}{12.1}  &&  \multirow{2}{*}{42.2} &  \multirow{2}{*}{} & \multirow{2}{*}{39.8}  &  \multirow{2}{*}{0.03}  &  \multirow{2}{*}{48.7}  &  \multirow{2}{*}{2.4}  \\
(diiodomethane-formamide-water) &  &  &  &   &&   &  &  &  \\
\hline
vOCG method &  \multirow{2}{*}{49.2} &  \multirow{2}{*}{} & \multirow{2}{*}{38.4} &  \multirow{2}{*}{0.62} &  \multirow{2}{*}{47.0} &  \multirow{2}{*}{10.7}  &&   \multirow{2}{*}{40.4} &  \multirow{2}{*}{}  & \multirow{2}{*}{39.8}  &  \multirow{2}{*}{0.002}  &  \multirow{2}{*}{50.5}  &  \multirow{2}{*}{0.6} \\
(diiodomethane-glycerol-water) &  &  &  &   &&   &  &  &  \\
\hline
vOCG method &  \multirow{2}{*}{n/a} &  \multirow{2}{*}{} & \multirow{2}{*}{n/a} &  \multirow{2}{*}{n/a} &  \multirow{2}{*}{n/a} &  \multirow{2}{*}{n/a}  &&   \multirow{2}{*}{40.4} &  \multirow{2}{*}{}  & \multirow{2}{*}{39.8}  &  \multirow{2}{*}{0.002}  &  \multirow{2}{*}{50.6}  &  \multirow{2}{*}{0.6} \\
(diiodomethane-DMSO-water) &  &  &  &   &&   &  &  &  \\
\bottomrule
\multicolumn{14}{c}{\footnotesize $^\ast$All units are in mJ/m$^2$.}
 \end{tabular}
 \end{table}

\subsection{The surface energy of tholin--direct force measurement}
Kwok and Neumann (1999) suggests that contact angle data should be disregarded if the solid surface dissolves or reacts with the liquid surface, because the liquid surface tension ($\gamma_l$) would change from the pure liquid. For our experiments, the plasma tholin appears to partially dissolve in all the polar test liquids, thus we measured the surface energy of plasma tholin with a different technique to avoid liquid dissolution. We used the SFA to directly measure the adhesion forces between two tholin surfaces in dry nitrogen to derive its solid surface energy.

Because the radius of the surfaces used in the SFA is much larger than the radius of the surfaces used in the AFM (R = 1.3 cm for SFA compared to R = 20 $\mu$m for the colloidal probe AFM, Yu et al., 2017a), the contact area achieved during the SFA experiments is around 2000 $\mu$m$^2$, much larger than the contact area achieved in AFM experiments ($<$ 1 $\mu$m$^2$). As a consequence, the SFA measurements are more susceptible to surface roughness and surface contaminants (e.g., McGuiggan et al., 2002, 2011). For the tholin coated surfaces used in the SFA, when the surfaces are first brought together, a short-range repulsive force was measured. The repulsive force begins approximately 180 nm from contact and no adhesive force is observed when the maximum load is approximately 1 mN. This behavior is generally observed when airborne contaminants are absorbed to the surface or when small particles are near the contact zone. However, on subsequent contacts, and at higher loads ($\sim$1.3 mN), the surfaces jump into an adhesive contact. During this jump, the contact area rapidly changes from $<$1 $\mu$m$^2$ to 30 $\mu$m$^2$ and increases with further loading. The film thickness is measured from this flattened adhesive contact position and the total surface separation was measured to be 76 nm, giving 38 nm film thickness per surface. The mica sheet thickness was measured to be 6.4 $\mu$m.

The adhesion force is measured by separating the surfaces. During the separation, the contact diameter decreases until the surfaces suddenly jump out of contact. The maximum pull-off force was measured to be 8.125 mN.

Using the JKR approximation, the pull-off force can be used to compute the surface energy ($\gamma_{sv}$) of the plasma tholin film:
\begin{equation}
 \gamma_{sv}=\frac{F_{\rm{pull-off}}}{3\pi R}.
\label{eq:jkr_results}
\end{equation}
For R = 1.3 cm and a pull-off force of 8.125 mN, the surface energy of plasma tholin is measured to be $\gamma_{sv}$ = 66 mJ/m$^2$. This is similar to what was derived from the OWRK two-liquid method (68.1$\pm$3.6 mJ/$m^2$). As we measured the contact angle immediately after the liquid is dropped on the surface, the liquid surface tension may not yet deviate too far before the dissolution happens.

Note that the surface energy calculated for layered surfaces, such as those measured here, may vary slightly from the formula as given in Equation \ref{eq:jkr_results} where homogenous surfaces are assumed. For a 38 nm tholin layer (Young's modulus for tholin around 10 GPa, Yu et al., 2018) on top of 6.4 $\mu$m mica (Young's modulus around 70 GPa) on top of a thick glue layer (Young's modulus around 3.4 GPa), the calculated surface energy will vary slightly depending on the amount of flattening. Therefore, the surface energy may be between 50 mJ/m$^2$ and 100 mJ/m$^2$ according to F$_{\rm{pull-off}}$/4$\pi$R $<$ $\gamma_{sv}$ $<$ F$_{\rm{pull-off}}$/2$\pi$R (McGuiggan et al., 2007).

\subsection{Predicting haze-clouds and haze-lakes interactions on Titan}
In this study, we only discuss the results assuming that the actual Titan aerosols are similar to tholin produced in our laboratory. In Titan's atmosphere, various organic species could condense to form clouds, and Titan's refractory haze particles could serve as the heterogenous cloud condensation nuclei (CCN) for these organic vapors (e.g., Anderson et al., 2018). Here we focus on studying two main cloud condensates, methane and ethane, both with adsorption and nucleation experiments performed on tholin made in another laboratory (Curtis et al., 2008). Heterogenous nucleation is the most efficient when the CCN are soluble in the cloud condensates according to the K\"ohler theory (K\"ohler, 1936), while Titan haze particles are likely insoluble in non-polar liquids including methane and ethane (e.g., Raulin, 1987; McKay, 1996). However, insoluble particles could also act as efficient CCN if they form a small contact angle with the cloud condensates species (McDonald, 1964; Mahata \& Alofs, 1975). Using the calculated surface energies of Titan tholins from Section 3.2 and 3.3, we can estimate the contact angles between tholins and species with known surface free energies to determine if Titan hazes can act as good CCN for methane and ethane clouds, both of which would condense from the gas phase to the solid phase to form ice clouds near the tropopause of Titan's atmosphere (Lavvas et al.,2011a).

The haze particles could also sediment down into the lakes and interact with the lake liquids. Titan hazes are likely insoluble in the dominant lake components (methane, ethane, and nitrogen, Mastrogiuseppe et al., 2016, 2018) as demonstrated by previous solubility experiments (McKay, 1996; Carrasco et al., 2009; Kawai et al., 2013; He \& Smith, 2014a). Thus, the haze particles can either sink in the Titan lakes and become lakebed sediments, or float on the surface of Titan's lakes. We need to consider both the Archimedes' buoyancy principle and the capillary process to determine the floatability of the Titan haze particles, as both could lead to the floatation of particles. The forces that act on an aerosol particle with radius $R$ that falls onto the surface of Titan lake liquids can be written as:
\begin{equation}
F_{tot}=2\pi R\gamma_l sin^2{\frac{\theta}{2}}-\frac{4}{3}\pi R^3(\rho_p-\rho_l)g,
\label{eq:balance}
\end{equation}
$\gamma_l$ is the surface tension of the lake liquids, $\theta$ is the contact angle between the aerosol particle and the lake liquids, $\rho_p$ is the density of the aerosol particle and $\rho_l$ is the density of the lake liquids. The first term on the right side of Equation \ref{eq:balance} is the maximum capillary force on the particle-liquid-air interface (the capillary process, Scheludko et al., 1976), and the second term is the buoyancy of the aerosol particle (Archimedes' buoyancy principle). Positive $F_{tot}$ means a total upward force, and the particle would float, and negative $F_{tot}$ means the particle would sink.

According to Archimedes' buoyancy principle, a material would float when the particle density is smaller than the liquid density ($\rho_p<\rho_l$), making the second term positive in Equation \ref{eq:balance}, and thus the total force always points upwards ($F_{tot}>0$ as the first term is always positive). Several groups have measured the densities of tholins produced with different experimental conditions and setups using different density measurement techniques, which varies between 500--1400 kg/m$^3$ (Imanaka et al., 2012; H\"orst \& Tolbert, 2013; He et al., 2017). The density of the Titan's lakes (a mixture of methane, ethane, and nitrogen) is estimated to be around 450--700 kg/m$^3$ (Cordier \& Carrasco, 2019). If the density of Titan haze particles is smaller than the density of the lake liquids, they would float on Titan's lakes. If the density of Titan haze particles is close to the higher end of the measured tholin density, we need to compare the buoyancy force to the capillary force; if the capillary force is smaller than the buoyancy force, the haze particles could sink to the bottom of the lakes; otherwise, they may float on Titan's lakes.

Capillary processes could also lead to the flotation of aerosol particles, when the capillary forces are larger than gravitational forces, the aerosol particles do not necessarily need lower density than the fluid to float. This process depends on the interfacial interaction between the solid aerosols and the liquid lakes species. Cordier \& Carrasco (2019) suggests that a material that is lyophobic to the fluid (with contact angle $\theta>90^\circ$) would be able to float, and the aerosols would form a floating layer of particles on Titan's lakes if they are lyophobic to the lake liquids. Thus, it is important to assess the contact angle between tholin and the Titan lake liquids to constrain the floatability of aerosols on Titan's surface.

We can use the surface free energies of two known materials and their corresponding surface free energy components to calculate the contact angle between them, by rewriting Equation \ref{eq:contactangle}:
\begin{equation}
cos\theta=\frac{2(\sqrt{\gamma^d_s\gamma^d_l}+\sqrt{\gamma^p_s\gamma^p_l})}{\gamma_l}-1,
\label{eq:contact_angle}
\end{equation}
this equation applies when $-1<$ cos$\theta<1$, and one material would make a finite contact angle on the other. When cos$\theta\geq$ 1, Equation \ref{eq:contact_angle} does not apply anymore, as once $\gamma^d_l\le\gamma^d_s$ and $\gamma^p_l\le\gamma^p_s$, the contact angle $\theta$ is zero, meaning that the liquid would completely wet the solid surface (or with one material spreads on the other, Fox \& Zisman, 1950, 1952, Gao \& Benneke, 2018; Powell et al., 2018).

Table \ref{table:hydrocarbons} summarizes the surface tensions of the dominant lake components in Titan's lakes, including methane, ethane, and nitrogen (Yaws \& Richmood, 2009). Using Table \ref{table:hydrocarbons} and the measured surface energies of tholins, we can calculate the contact angles between each lake component and tholin, the results are summarized in Table \ref{table:prediction}. Tholin would be completely wetted by Titan's lake liquids (or $\theta=0$), as the dispersive surface tensions ($\gamma^d_l$) of all the lake liquids are smaller than the dispersive surface energies of both tholins ($\gamma^d_s$), even considering the standard deviations of the surface energy measurements, and these liquids are all non-polar with zero polar components, so $\gamma^p_l<\gamma^p_s$. When $\theta=0$, the first term of Equation \ref{eq:balance} would equal to zero, and the total force $F_{tot}$ would always point downwards if the density of the aerosols is higher than the density of the lake liquids. In this case, it is unlikely that the Titan hazes would float on the lakes. Note that the tholin samples we used are quite smooth (with RMS roughness of $<$ 5 nm), while the roughness of the actual aerosols could be much higher than the laboratory produced tholin, which could affect the contact angle/wettability of the aerosols. There are two theoretical models that describe the effect of roughness on the apparent contact angle ($\theta_a$) for rough surfaces. The Wenzel model (Wenzel, 1936) applies to both lyophilic and lyophobic surfaces, and predicts that a lyophobic surface would behave more lyophilic with surface roughness, while a lyophobic surface would become more lyophobic with surface roughness:
\begin{equation}
cos\theta_a=rcos\theta,
\end{equation}  
$r$ is the roughness factor, which is defined as the ratio between the total surface area of the rough surface over the apparent surface area (the projected area of the rough surface), for a smooth surface, $r=1$; and $r$ is always larger than 1 for rough surfaces, $\theta$ is the contact angle for a smooth surface. The Cassie-Baxer model (Cassie \& Baxer, 1944) typically applies to lyophobic surfaces ($\theta>90^\circ$), when the liquid cannot penetrate the surface asperities, and it also predicts a more lyophobic surface with higher surface roughness. The contact angles between Titan's dominant lake components and tholin are all zero, where the Wenzel model applies, thus the addition of surface roughness would keep the aerosol to be lyophilic to all the lakes components ($\theta=0$).

Laboratory produced tholins are considered decent analogs to the haze particles in Titan's atmosphere as they at least partially match both the Huygens Aerosol Collector Pyrolyzer (ACP) measurements (Isra\"el et al., 2005; Coll et al., 2013), the Cassini Plasma Spectrometer (CAPS) measurements (Sciamma-O'Brien et al., 2014; Dubois et al., 2019), the Cassini Ion and Neutral Mass Spectrometer (INMS) measurements (Dubois et al., 2020), and the Cassini Visible and Infrared Mapping Spectrometer (VIMS) spectra (Gautier et al., 2012; Sciamma-O'Brien et al., 2017). Laboratory made tholins may have different chemical and physical properties compared to real Titan aerosol particles, as the energy sources used in these experiments (cold plasma or UV) cannot fully resemble the energy sources for aerosol production in Titan's atmosphere. Titan's haze particles could also be coated with condensable species in the atmosphere before they fall to the lakes. It is possible that the actual Titan aerosols might be different from the measured tholin samples. In this case, if the non-polar component of the surface energy of the actual Titan aerosols is less than 20-30 mJ/m$^2$, based on Equation \ref{eq:contact_angle}, the contact angle would be larger than 0, then small aerosols particles would be able to float on Titan's lakes (Scheludko et al., 1976).

As the aerosol particles would likely sink into Titan's lakes as lakebed sediments, there is another application for our predicted contact angles between tholin and Titan's lakes liquids. Previous works have suggested the formation of gas bubbles due to nitrogen exsolution in Titan lakes (Cordier et al., 2017; Malaska et al., 2017; Cordier \& Liger-Belair, 2018; Farnworth et al., 2019), which may explain the transient features observed by the Radio Detection and Ranging (RADAR) instrument onboard Cassini on the surface of lakes (Hofgartner et al., 2014, 2016). The ascending bubbles could also potentially lead to flotation of sediments from Titan's lakebed. According to the bubble flotation theory and experiments, flotability of sediments with the aid of ascending bubble disappears when contact angle between the sediments and the liquid is smaller than $\sim$10--20$^\circ$ (Scheludko et al., 1976; Crawford \& Ralston, 1988). Thus, if the aerosols sink into the lakes and become lakebed sediments and are similar in composition to tholin, it is unlikely they will be brought up by the nitrogen bubbles because the contact angle between tholin and Titan's lake liquids is zero, which is well below the contact angle threshold. Titan's lakebed sediments could also be composed of other substances such as eroded crustal materials transported by rivers, which may be able to float on Titan's lakes.

Table \ref{table:hydrocarbons} also summarizes the surface energies of the main ice clouds in Titan's atmosphere, methane and ethane (Guez et al., 1997; Lavvas et al., 2011a), whose surface energies are estimated based on liquid phase relaxation using the ratio of the latent heats of sublimation and vaporization (Pruppacher \& Klett, 1978). We can use Table \ref{table:hydrocarbons} and the surface energies of tholins (OWRK two-liquid method derivation results) to calculate the contact angle between the ice clouds and tholin. The results are summarized in Table \ref{table:prediction2}. Overall, methane ice would form a zero contact angle on tholin, no matter which surface energy value of tholin is used from Table \ref{table:surfaceenergy}. Thus, the Titan haze particles, if they are similar to tholins, should be good CCN for methane ice clouds. Ethane ice would have a 20--30$^\circ$ angle on plasma tholin and a 15--22$^\circ$ angle on UV tholin using the surface energy of tholin derived using the OWRK two-liquid method. The surface energy values derived from OWRK two-liquid method, vOCG method, and the direct force measurement conforms to the literature values derived from ethane adsorption experiments performed with tholin made in a different laboratory (Barth \& Toon, 2006; Curtis et al., 2008; Lavvas et al., 2011a; Rannou et al., 2019). However, the contact angle of ethane ice would be 55.7$^\circ$ for plasma tholin and 54.1$^\circ$ for UV tholin if using surface energy derived from OWRK-fit method, which is very different from the results derived with the OWRK two-liquid method. As discussed in Section 3.2, the OWRK-fit method likley underestimates the surface energy of tholin as many test liquids have lower surface tension components compared to tholin. Thus, here we will adopt the contact angle estimation of 15--30$^\circ$ for ethane on tholin surfaces. For insoluble particles to act as efficient CCN, McDonald (1964) gave a contact angle upper limit of $12^\circ$, while Mahata \& Alofs (1975) found a higher upper limit of $\sim30^\circ$ for rough particles. The contact angle between ethane ice and tholins roughly fits in this regime and thus the Titan haze particles are likely good CCN for ethane ice clouds as well.

\begin{table}
\caption{The surface tensions of Titan's dominant lake components (methane, ethane, and nitrogen) and the surface energies of the main Titan ice cloud condensates (methane and ethane). We use the surface tension values of liquid methane, ethane, and nitrogen and their corresponding surface tension components at Titan's surface conditions (1.5 bar and 94 K, Yaws \& Richmood, 2009). Unit is mN/m for all surface tension numbers in the table. $\gamma_l$ is the total surface tension. $\gamma_l^{d}$ and $\gamma_l^{p}$ are respectively the dispersive and polar components. The surface energies of solid methane and ethane are approximated as $\gamma_s=(L_{sub}/L_{vap})^2\gamma_l$ (Guez et al., 1997), where $L_{sub}$ and $L_{vap}$ are the enthalpy of sublimation and vaporization at the triple point of the material (Guez et al., 1997). For methane, $L_{sub}=9.7$ kJ/mol and $L_{vap}=$ 8.6 kJ/mol. For ethane, $L_{sub}=20.5$ kJ/mol and $L_{vap}=$ 17.7 kJ/mol (Bondi, 1963; Stephenson \& Malanowski, 1987).}
 \label{table:hydrocarbons}
 \centering
 \begin{tabular}{c  c  c c r crcc}\toprule
 Species & \multicolumn{3}{c}{Liquid at 94 K} && \multicolumn{4}{c}{Solid $<$ 90 K}\\
 \cmidrule{2-4}  \cmidrule{6-9}
 & Total surface tension$^\ast$  & \multicolumn{2}{c}{OWRK components$^\ast$}  && Total surface energy$^\dag$  && \multicolumn{2}{c}{OWRK components$^\dag$}\\
 & $\gamma_l$ & $\gamma_l^d$ & $\gamma_l^p$  && $\gamma_s$ && $\gamma_s^d$ & $\gamma_s^p$\\
\midrule
Methane & 17.8 & 17.8 & 0 && 23.5 && 23.5 & 0\\
Ethane & 31.8 & 31.8 & 0 && 43.4 && 43.4 & 0 \\
Nitrogen & 5.3 & 5.3 & 0 && n/a && n/a & n/a\\
\bottomrule
\multicolumn{9}{c}{\footnotesize$^\ast$ Units are in mN/m. $^\dag$ Units are in mJ/m$^2$.}
 \end{tabular}
 \end{table}

\begin{table}
\caption{The predicted contact angles between dominant liquid components in Titan's lakes and tholins (plasma and UV) using the surface tensions of the liquids from Table \ref{table:hydrocarbons} and the surface energies of tholins derived by the OWRK two-liquid method in Table \ref{table:surfaceenergy}.}
 \label{table:prediction}
 \centering
 \hspace*{-1.5cm}
\begin{tabular}{c  ccrcc}\toprule
 Liquid &  \multicolumn{5}{c}{Predicted contact angle} \\
 \cmidrule{2-6}
 &  \multicolumn{2}{c}{Plasma Tholin} && \multicolumn{2}{c}{UV Tholin} \\
  \cmidrule{2-3}   \cmidrule{5-6} 
 & cos$\theta$ & $\theta$  && cos$\theta$ & $\theta$ \\
\midrule
Methane & 1.98$\pm$0.05 ($>$1) & 0$^\circ$ && 2.04$\pm$0.03 ($>$1)  & 0$^\circ$ \\
Ethane & 1.23$\pm$0.040 ($>$1) & 0$^\circ$ && 1.27$\pm$0.03 ($>$1) & 0$^\circ$  \\
Nitrogen & 4.47$\pm$0.10 ($>$1) & 0$^\circ$ && 4.56$\pm$0.06 ($>$1)  & 0$^\circ$ \\
\bottomrule
 \end{tabular}
 \end{table}
 
 \begin{table}
\caption{The predicted contact angles between solid condensates (methane and ethane) in Titan's atmosphere and tholins (plasma and UV) using the surface energy of the solids from Table \ref{table:hydrocarbons} and the surface energies of tholins derived by the OWRK two-liquid method in Table \ref{table:surfaceenergy}. The literature values of the contact angle are derived from methane and ethane adsorption experiments performed by Curtis et al., (2008) using tholin produced in a different laboratory. The corresponding citations are: [a] Rannou et al., 2019; [b] Barth \& Toon, 2006; [c] Lavvas et al., 2011a.}
 \label{table:prediction2}
\centering
\hspace*{-4.2cm}
\resizebox{9in}{!}{
\begin{tabular}{c  ccrcc r c c rccrcc}\toprule
 Solid &  \multicolumn{5}{c}{Predicted contact angle} && \multicolumn{8}{c}{Literature values} \\
 \cmidrule{2-6} \cmidrule{8-15}
 &  \multicolumn{2}{c}{Plasma Tholin} && \multicolumn{2}{c}{UV Tholin} && \multicolumn{2}{c}{[a]} &&  \multicolumn{2}{c}{[b]} &&  \multicolumn{2}{c}{[c]} \\
 \cmidrule{2-3}  \cmidrule{5-6} \cmidrule{8-9} \cmidrule{11-12} \cmidrule{14-15}
 & cos$\theta$ & $\theta$  && cos$\theta$ & $\theta$ && cos$\theta$ & $\theta$  && cos$\theta$ & $\theta$ && cos$\theta$ & $\theta$\\
\midrule
Methane & 1.596$\pm$0.046 ($>$1) & 0$^\circ$ && 1.645$\pm$0.029 ($>$1)  & 0$^\circ$ && 0.994$\pm$0.001 & 5.7--6.8$^\circ$ && 0.981 & 11.2$^\circ$ && 0.995 & 5.7$^\circ$\\
Ethane & 0.910$\pm$0.033 & 19.2--28.8$^\circ$ && 0.946$\pm$0.021 & 14.7--22.4$^\circ$ && 0.966$\pm$0.007 & 13.3--16.5$^\circ$ && 0.986 & 9.6$^\circ$ && 0.979 & 11.8$^\circ$\\
\bottomrule
 \end{tabular}}
 \end{table}
 
\section{Discussion and Conclusion}
To better understand the physical processes involving the solid haze particles on Titan, we quantified an important physical property of the haze, the surface energy. We measured the contact angles between the Titan haze analogs (``tholin"), and a variety of liquids with different surface tensions to estimate its surface energy. We found that the tholins produced by two distinct energy sources (cold plasma and UV irradiation) have relatively similar total surface energies, which indicates that they may have similar cohesive properties. Both tholins have significant polar components, indicating the abundance of polar structures on the surface of the material (e.g., Quirico et al., 2008; Derenne et al., 2012; He et al., 2012; He \& Smith, 2014b; Maillard et al., 2020). The polar component of the UV tholin is smaller than cold plasma tholin, which could be caused by less efficient incorporation of nitrogen in the final solid material, as the UV lamp used in our study cannot directly dissociate nitrogen in the gas mixture. The total surface energy of tholin derived by the OWRK and Wu two-liquid component methods yields $\sim$60-70 mJ/m$^2$. This value can be used to cross-compare with literature values because the two-liquid component methods are the most widely used derivation methods. The OWRK-fit method incorporates the most data from multiple test liquids and the surface energy is calculated to be $\sim$50--55 mJ/m$^2$, however, this method may underestimate the total surface energy of tholin as several low surface tension liquids used may not be able to fully probe the dispersive/polar properties of tholin. We also used the direct force measurement to validate the contact angle measurement for plasma tholin, which dissolves partially in most polar test liquids. The direct force measurement yields a surface energy value of plasma tholin of 66 mJ/m$^2$, which is similar to the results derived from the contact angle measurement using OWRK two-liquid method.

Because the tholin surfaces were synthesized in an ultra-high vacuum (UHV) chamber with N$_2$ and CH$_4$ gas mixture, exposing the films to the ambient atmosphere during measurements may affect the surfaces. For example, the tholin films could be chemically altered upon exposure to oxygen and water. In addition, water moisture from the atmosphere as well as airborne contaminants may adsorb on the surfaces. All these occurrences could affect the surface properties of the tholin samples. The samples were stored in an oxygen and moisture free nitrogen glove box to minimize chemical alteration and adsorption prior to the measurements. The contact angle measurement were performed in ambient air within two hours after we removed the surfaces from the glove box. Thus, chemical alternation by oxygen and water should not be significant as these processes occur on a longer time scale ($\sim$days, e.g., Cable et al., 2012). On the other hand, liquid adsorption usually occurs in a short timescale ($\sim$minutes, Yu et al., 2017b), thus adsorption likely occurred during the contact angle measurement. Adsorption of water moisture on a solid could decrease the surface energy of a high surface solid surface, and the effect is most prominent for high energy surfaces such as metals and silicates, with surface energies of hundreds to thousands mJ/m$^2$ (Wu, 1982). However, the effect is less pronounced for solids with similar or lower surface energies than water. The direct force measurement, which is conducted in dry nitrogen, yield surface energy value of plasma tholin similar to the surface tension of water, thus, tholin is probably not a high energy surface like silicates or metals, and the adsorption effect of water in ambient air should be less pronounced. Airborne containments like hydrocarbons could also adsorb onto the tholin surfaces. Hydrocarbons are known to adsorb onto surfaces to decrease the surface free energy. Because the measured surface energy of tholin is higher than the surface tension of hydrocarbons, the adsorption of hydrocarbons should decrease the surface energy, as previously mentioned. Thus, we consider the effect of hydrocarbons or vapors adsorption to give rise to a smaller surface energy than would be measured in vacuum and therefore the measured surface energy should be a lower limit.  In spite of the likely presence of an adsorbed layer on the tholin surfaces, the measured surface energy values are still quite high for a surface comprised of hydrogen, carbon, and nitrogen, indicating that tholin is a highly cohesive substance.

We found that plasma tholin would partially dissolve in all the polar solvents used in the contact angle measurement. As seen in Figure \ref{fig:time}, the contact angle slowly decreased with time, most notably for the plasma tholin. According to the Young-Dupr\'e equation (Equation \ref{eq:youngdupre2}), if the contact angle decreases (cos$\theta$ increases), then either $\gamma_s-\gamma_{sl}$ increases, or $\gamma_l$ decreases, or both. Assume $\gamma_s-\gamma_{sl}$ remains constant, then $\gamma_l$ must decrease, indicating that the dissolved plasma tholin could change the surface tension of the test liquid. Even though the drop profile was captured as soon as the liquids wet the solid surface, some dissolved substances surface active materials could significantly decrease the surface tension of the liquid to affect the interpretation of the contact angle measurement (van Oss, 2006). To further investigate the effect of solubility on the test liquids' surface tension, we use the pendent drop method to measure the surface tension of water before and after it dissolves plasma tholin. The measurement was performed using an Ossila optical tensiometer, which captures the shape of a droplet suspended at the end of a 25 $\mu$L syringe needle. We measured the drop shapes of both pure deionized water and the same water with dissolved plasma tholin, which were then fitted with the software algorithm to determine its surface tension. We added $\sim$5 mg of plasma tholin to $\sim$1 ml of water for $\sim$1 hour and measured the surface tension of the supernatant, and found a 5--10 mN/m surface tension decrease compared to pure water. Using the maximum measured surface tension change of 10 mN/m (giving a $\gamma_l$ = 62.8 mN/m), the range of the surface energy of tholin (using OWRK two-liquid method with diiodomethane) is calculated to be between 60--66 mJ/m$^2$ (the surface tension change could affect either the polar or the dispersive components of water), which does not deviate too much from the original results ($\sim$68 mJ/m$^2$). Since the droplet shape was captured in less than 5 seconds and the solution is much more diluted, the surface tension change of water should be less than 5-10 mN/m and thus likely has minimal effect on the tholin surface energy calculation.

Overall, tholins have relatively high surface energies compared to common polymers, which supports its high cohesiveness as measured by colloidal probe AFM (Yu et al., 2017a). As discussed in Yu et al., (2017a), we do not expect the surface energy of tholin to vary much if extrapolated to Titan's surface temperature (94 K), as tholin's melting point is at least 50 $^\circ$C above the temperature where the measurements were made (He \& Smith, 2014a). We estimated the amount of change with temperature (94 K and 300 K) according to the Lifshitz theory (Lifshitz \& Hamermesh, 1992), and we found a less than 5\% decrease in the total surface energy. So coagulation of Titan haze particles could be efficient once they collide within each other. If the surface sand is similar to tholin, the sand particles could be highly cohesive and thus would need higher wind speed to be mobilized.

\begin{figure}[h]
\centering
\includegraphics[width=\textwidth]{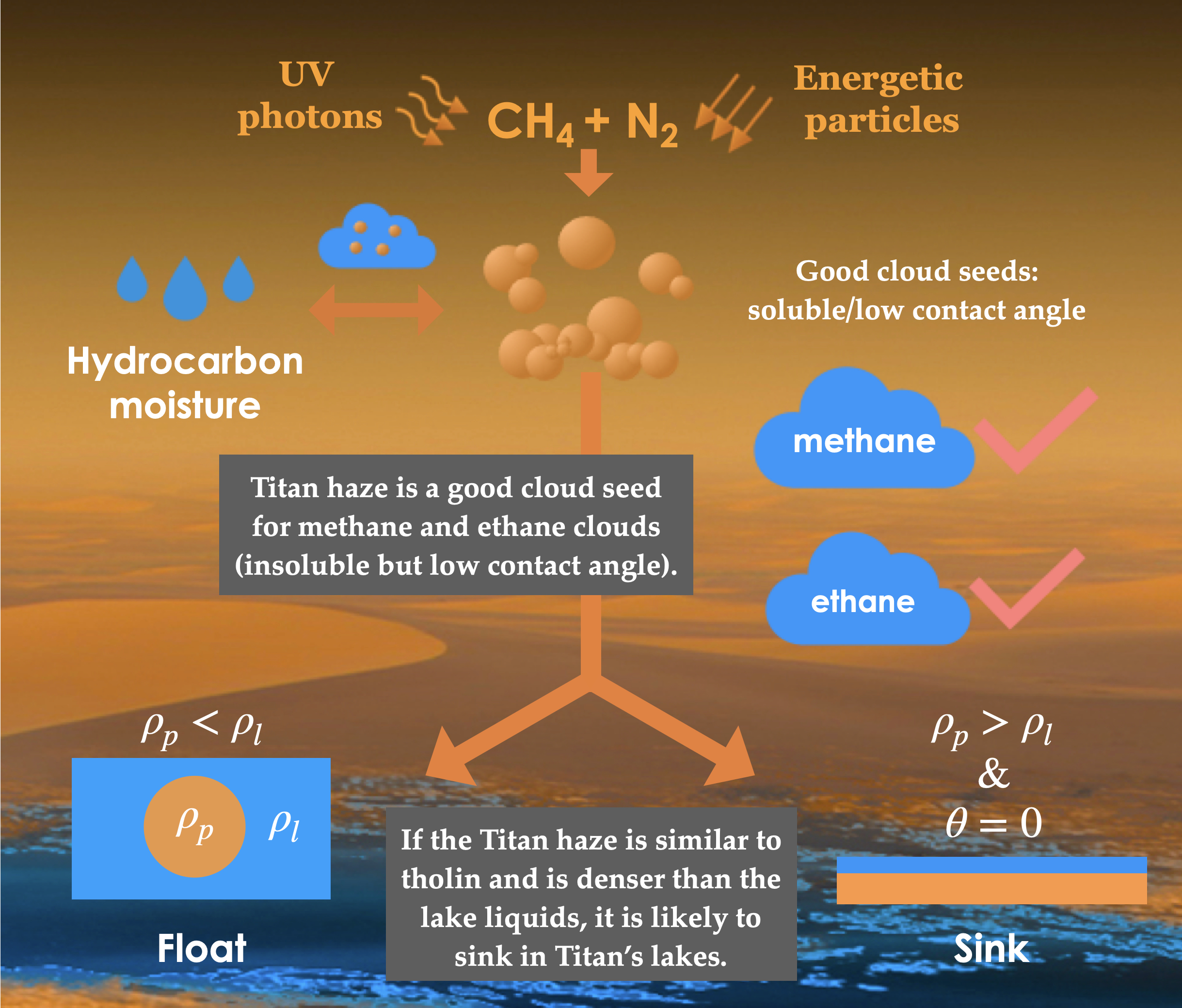}
\caption{Titan's haze particles are likely probably good cloud condensation nuclei (CCN) to heterogeneously nucleate methane and ethane ice clouds in Titan's atmosphere because the contact angles predicted between tholin and solid methane/ethane ices are small. When the haze particles fall unto Titan's lakes, if the density of the haze particles ($\rho_p$) is smaller than the lake liquids ($\rho_l$), they would float on Titan's lakes. If the density of the haze particles is larger than the lake liquids ($\rho_p>\rho_l$), and if the Titan haze is similar to tholin, they are unlikely to float on the surface of the lakes and dampen the surface waves as the contact angle between the lake liquids and tholin is zero ($\theta=0$). In this case, they would sink into the lakes becoming lake bed sediments.}
\label{fig:summary}
\end{figure}

The measured surface energy of tholin can also be used together with the wetting theory to estimate the wetting scheme between tholin and other substances, which could inform us on heterogeneous cloud nucleation efficiency and aerosol-lake interactions. The refractory aerosols could act as cloud condensation nuclei (CCN) for condensates species in Titan's atmosphere and allow heterogeneous nucleation and efficient cloud growth. The best cloud condensation nuclei for a condensable liquid are the soluble species (Volmer, 1939). Both UV and plasma tholins are insoluble in non-polar solvents (e.g., He \& Smith, 2014a). But materials with low contact angles ($<30^\circ$) can also serve as CCN for heterogenous nucleation (Mahata \& Alofs, 1975), thus Titan haze could still be good CCN to form the main hydrocarbon ice clouds (methane and ethane) in Titan's atmosphere, because the contact angles between tholin and methane/ethane ices are relatively low (Figure \ref{fig:summary}). 

When the aerosol particles fall unto the surface of the Titan's lakes, the aerosols could either float or sink into the lake liquids. Flotable aerosols could contribute to surface wave dampening which may explain the nearly waveless lakes observed by Cassini (Cordier \& Carrasco, 2019). If the aerosols sink instead, they will become lakebed sediments. To determine whether the aerosols float or sink, two properties of the aerosols need to be taken into account, their density and the contact angle formed between the aerosols and the lake liquids. If the density of the aerosols ($\rho_p$) is smaller than the density of the lake liquids ($\rho_l$), the aerosols would float on the lakes because of buoyancy (Figure \ref{fig:summary}). If the density of the aerosols is higher than the lake liquids ($\rho_p>\rho_l$), we need to assess the contact angle between the aerosols and the lake liquids to determine the capillary force could lead to flotation of the particles. Since the dispersive component of tholin's surface energy is larger than the surface tensions of Titan's dominant non-polar lake liquids (methane, ethane, and nitrogen), the lake liquids would completely wet tholin, with a contact angle $\theta=0$, leading to a zero capillary force. Thus, if $\rho_p>\rho_l$, because the contact angle formed between tholin and the lake liquids is zero (or zero capillary force), the aerosols would sink into Titan's lakes.

It is possible that the Titan haze particles are coated with condensable species in the atmosphere before they fall unto the lakes (Figure \ref{fig:summary}). Most of the condensates (CH$_4$, C$_2$H$_4$, C$_2$H$_6$) will be lost through photochemistry or converted to liquid phases before they fall unto the surface (Anderson et al., 2018). The condensation and melting processes could change the macrostructure of the aerosols by compacting and smoothing the particles (e.g., Lavvas et al., 2011b; Ma et al., 2013), but likely will not change the chemical structure of the aerosols as laboratory produced tholins are insoluble in non-polar solvents (McKay, 1996; Carrasco et al., 2009; Kawai et al., 2014; He \& Smith, 2014a). Other condensable species, such as HCN, C$_6$H$_6$, HC$_3$N, would remain as solid ices when they fall to the surface, and a more detailed study of ice interactions with Titan lakes is currently ongoing. The aerosols deposited on the surface or on the bottom of the lakes may also be altered highly-penetrative irradiation such as galactic cosmic rays (GCR), which could change their chemical structure (Sittler et al., 2020), and future laboratory study with irradiated tholins could inform on the surface properties change of the aerosols. Depending on the density of the Titan haze particles compared to the lake liquids, tholin particles could float on Titan's lakes but the surface of the particles will be coated by the lake liquids.  

\acknowledgments
X. Yu is supported by the 51 Pegasi b Fellowship from the Heising-Simons Foundation. P. McGuiggan is supported by the 3M Nontenured Faculty Grant. X. Zhang is supported by NASA Solar System Workings Grant 80NSSC19K0791. We thank Xu Yang for technical support in R programming. We are also grateful for Sam Birch and Shannon Mackenzie for their valuable inputs on Cassini observations relevant to this work.

\end{document}